\documentclass[a4paper]{iopart}

\usepackage[english]{babel}
\usepackage[dvips]{graphicx}

\usepackage{amssymb}
\usepackage{amsthm}
\usepackage{bm}
\usepackage{cite}
\usepackage{color}
\usepackage{iopams}
\usepackage{latexsym}
\usepackage{times}
\usepackage{url}

\usepackage{hyperref}
\hypersetup{
  colorlinks=true,        
  linkcolor=blue,         
  citecolor=cyan,         
}

\newcommand{\mss}{\,{\rm ms}}
\newcommand{\km}{\,{\mathrm{km}}}
\newcommand{\gcm}{\,{\mathrm{g}/\mathrm{cm}^{3}}}
\newcommand{\Msun}{M_{\odot}}

\newcommand{\cf}{cf.~}
\newcommand{\ie}{i.e.,~}
\newcommand{\eg}{e.g.,~}

\begin{document}

\title{On the use of tracer particles in simulations of binary neutron stars} 
\author{Luke Bovard$^{1}$ and Luciano Rezzolla$^{1,2}$}
\address{$^{1}$Institute for Theoretical Physics, Max-von-Laue-Str. 1,
  60438 Frankfurt, Germany\\
$^{2}$Frankfurt Institute for Advanced Studies, Ruth-Moufang-Str. 1,
  60438 Frankfurt, Germany}

\date{\today}
\begin{abstract}
In grid-based codes that provide the combined solution of the Einstein
equations and of relativistic hydrodynamics, the history of the fluid is
not simple to track, especially when compared with particle-based codes.
The use of tracers, namely massless particles that are advected with the
flow, represents a simple and effective way to solve this problem. Yet,
the use of tracers in numerical relativity is far from being settled and
several issues, such as the impact of different placements in time and
space of the tracers, or the relation between the placement and the
description of the underlying fluid, have not yet been addressed. In this
paper we present the first detailed discussion of the use tracers in
numerical-relativity simulations focussing on both unbound material --
such as the one needed as input for nuclear-reaction networks calculating
r-process nucleosynthesis in mergers of neutron stars -- and on bound
material -- such as the one in the core of the object produced from the
merger of two neutron stars. In particular, when interested in unbound
matter, we have evaluated different placement schemes that could be used
to initially distribute the tracers and how well their predictions match
those obtained when using information from the actual fluid flow.
Countering our naive expectations, we found that the most effective
method does not rely on the rest-mass density distribution nor on the
fluid that is unbound, but simply distributes tracers uniformly in
rest-mass density. This prescription leads to the closest matching with
the information obtained from the hydrodynamical solution. When
considering bound matter, we demonstrate that tracers can provide insight
into the fine details of the fluid motion as they can be used to track
the evolution of fluid elements or to calculate the variation of
quantities that are conserved along streamlines of adiabatic flows.
\end{abstract}

\maketitle

\section{Introduction}

The recent detection of gravitational waves from binary black hole
mergers \cite{Abbott2016a} by LIGO has heralded the era of
gravitational-wave astronomy. Detectors such as Virgo, KAGRA, and the
Einstein Telescope (ET) \cite{Accadia2011_etal, Kuroda2010, Punturo:2010}
are expected to detect further events that will allow for a completely
new way of observing the universe. An equally exciting possibility is the
detection of an electromagnetic counterpart simultaneous with a
gravitational wave detection which could help explain the long-standing
puzzle of short-gamma ray bursts \cite{Narayan92, Eichler89,
  Rezzolla:2011,Bartos:2012, Berger2013b}. Although only black hole
mergers have so far been detected, binary neutron star (BNS) mergers are
expected to be observed in the coming years. As such, significant
progress has been made over the last decade to accurately simulate their
merger (see, \eg \cite{Baiotti2016} for a recent review).

A recent area that has received significant interest is the possibility
that BNS mergers are the most likely progenitors of the heavy elements in
the universe \cite{Lattimer74, Li1998, Tanvir2013, Berger2013,
Tanaka2013, Rosswog2014a, Sekiguchi2015, Radice2016}. During the merger
process, the binary mergers ejects an amount of material that is on the
order of $10^{-2} - 10^{-4}\,\Msun$ \cite{Hotokezaka2013, Bauswein2013b,
Radice2016, Dietrich2016} and this material exists in an environment that
is very conducive towards rapid process neutron capture, r-process,
nucleosynthesis \cite{Li1998, Rosswog05, Arnould2007, Metzger:2010,
Roberts2011}. Due to a neutron rich environment required for the
reactions to occur, the potential astrophysical sites that could produce
the r-process elements are limited. For many years, it was thought that
core collapse supernova (CCSN) were the source of the heavy elements in
our universe \cite{Rosswog2015, Martin2015a}. However, recent simulations
of CCSN have shown that although CCSN eject sufficient amounts of
material, it does not appear to be neutron-rich enough to agree with
observations of the abundances observed in the solar system
\cite{Arcones2013a}. Furthermore, experimental observations of
metallicity in dwarf galaxies disfavours CCSN and instead point towards
BNS mergers as being the primary source of r-process \cite{Ji2016}.
Likewise, recent numerical simulations of BNS mergers with improved
neutrino microphysics support the BNS merger scenario as the ejecta from
the mergers is very neutron rich \cite{Baiotti2016}. Furthermore, as this
ejected material expands, it undergoes radioactive decay and could
potentially create an electromagnetic counterpart, the so-called
``kilonova'' \cite{Li1998, Tanaka2016, Metzger2017}. Hence, there is an
increasing interest in studying the composition, evolution and outflow
from a BNS merger and the different types of mechanisms that can cause
the fluid to undergo nucleosynthesis. There are four types of outflow
from a merger that have been considered: dynamical ejecta
\cite{Rezzolla:2010, Roberts2011, Kyutoku2012, Rosswog2013a,
Bauswein2013b}, neutrino-driven winds \cite{Dessart2009, Perego2014,
Wanajo2014, Just2015, Sekiguchi2015, Martin2015, Martin2015a, Just2016,
Sekiguchi2016, Murguia-Berthier2016}, magnetically driven winds
\cite{Shibata2011b, Kiuchi2012b, Siegel2014, Rezzolla2014b, Ciolfi2014},
and viscous evolution of the accretion disc \cite{Goriely2011,
Kastaun2014}. All these ejection channels have different ejection
properties which could potentially produce different nucleosynthetic
signatures and different light curves emitted from a kilonova,
\cite{Rosswog2013, Piran2013, Grossman2014, Perego2014, Wanajo2014,
Just2015, Sekiguchi2015, Radice2016, Just2016,
Sekiguchi2016,Tanaka2016,Metzger2017}.

In order to study the nucleosynthesis produced from these different
categories of ejecta, the thermodynamic history \eg the temperature,
entropy, and electron fraction, of a fluid element is required. These
time series are then used as the input for nuclear-reaction networks to
determine the abundances of the different elements \cite{Martin2015a}.
However, the timescales required for the nuclear reactions to occur, on
the order of seconds to days, is far beyond the current capabilities of
numerical-relativity codes, which can run at most a few tens of
milliseconds after the merger. In the ideal case, a nuclear-reaction
network would be solved simultaneously with the fluid evolution in the
simulation, however currently this is numerically infeasible and all
nuclear reactions must be treated in a post-processing step. Furthermore,
the different ejecta mechanisms have different timescales over which they
dominate, ranging from $\sim 10\mss$ for the dynamical ejecta, to $\sim 1
\, \mathrm{s}$ for the viscous ejecta in the disk. This difference in
timescales requires different numerical models and assumptions to be made
and to overcome these technical issues and various approaches have been
used by numerous groups.

One approach is to use Newtonian codes, eschewing general relativity
completely \cite{Fernandez2013,Martin2015a}. An advantage of Newtonian
codes is that they are computationally and mathematically much simpler
and can be run for longer times. Hence, they have been used to study the
long-term evolution of the fluid as it outflows, such as that required by
the neutrino driven winds -- which has timescales of $\sim 100\mss$
\cite{Martin2015a} -- and the viscous forces of the accretion disk
\cite{Fernandez2013}. These codes also study the nucleosynthesis produced
from the ejecta and the fluid history is recorded by tracer particles
(hereafter simply ``tracers''). Another approach is to use smoothed
particle hydrodynamics (SPH) \cite{Piran2013, Rosswog2014a, Grossman2014,
Metzger2015, Martin2015a} instead of grid-based codes. One major
advantage here is that it allows for exact evolution of the fluid's
thermodynamic properties because the fluid itself is made of interacting
particles. While this approach is beneficial for the nucleosynthesis
analysis, normally it does not provide solutions in full general
relativity. Furthermore this approach cannot be evolved for the long term
as the Newtonian codes can. Finally, for grid-based codes which fully
solve the Einstein equations, there are two approaches that have been
used to follow the thermodynamic history of the ejecta. The first is that
of tracers which passively follow the fluid as it evolves recording the
properties of the fluid as it evolves \cite{Wanajo2014}, providing a
history of the fluid lines. The second is to use a spherical surface that
the fluid passes through and use that as the initial conditions for the
ejecta; assuming the fluid to be undergoing adiabatic expansion, it is
then possible to extrapolate the dynamics to long timescales
\cite{Radice2016}. While both methods can successfully reproduce the
r-process, \cite{Wanajo2014, Radice2016, Bovard2017}, there is a benefit
in using the former over the latter. The second method, in fact, is
simpler to implement but it does not allow for the history of the fluid
element to be recorded. With the thermodynamical history of the fluid
element, the tracers are able to be input into radiative-transfer codes
that allow for the calculation of kilonova light curves
\cite{Grossman2014}. Thus, it is important that the tracers accurately
represent the underlying fluid.

Besides the study of the ejected matter, another area of research in
which the use of tracers is particularly beneficial is the evolution of
the binary merger product (BMP), that is, of the metastable object that
forms after a binary neutron merger. Significant work recently has gone
into understanding the nature of the stability of this BMP
\cite{Baiotti2016} and tracers provide a novel way of interpreting the
resulting behaviour of the BMP that is otherwise inaccessible to only
studying the fluid evolution \cite{Hanauske2016, Alford2017}.

Although simple in principle, the use of tracers in numerical relativity
is far from settled. As observational quantities can be computed from
tracer data, it is critical to ensure that the tracers used accurately
represent the underlying fluid. As such, many issues such as the effects
of placements of the tracers and of assigning tracers mass have not been
adequately discussed or settled. Furthermore, tracers are an inherently
particle-based idea while the most advanced GR simulations use a
grid-based code. Thus relating how tracers properly relate to the
underlying fluid requires special care which we discuss in detail. In
this paper we present the first detailed discussion of the use tracers in
numerical-relativity simulations. We focus on two areas: \emph{unbound
material} in the form of dynamical ejecta, and \emph{bound material} in
the core of a BMP formed from the merger of two neutron stars. More
specifically, when considering unbound matter, we have evaluated four
different placement schemes that could be used to initially distribute
the tracers and how well their predictions match those obtained when
using information from the actual fluid flow. Contradicting our naive
expectations that the best placements are those that are correlated with
the rest-mass density distribution or that follow the fluid that is
marked unbound, we have found that the most effective method is to
\emph{uniformly} sample across the rest-mass density distribution as this
leads to the closest matching with the hydrodynamical information on the
unbound material flowing across the computational domain. On the other
hand, when considering bound matter, we have shown that tracers can
provide insight into the stability of the BMP as they can be used to
track the evolution of fluid elements or to calculate the evolution of
quantities conserved along streamlines that would otherwise be
inaccessible in a grid-based code \cite{Hanauske2016, Kastaun2016}.

The paper is organised as follows: in Sec. \ref{sec:mans} we review the
mathematical and numerical setup employed to solve the equations of
relativistic hydrodynamics but also those describing the motion of the
tracers and their analysis. Section \ref{sec:unbound} is instead
dedicated to the study of the dynamics of unbound material and to the
discussion of the various placement schemes that we have considered. The
results in this section should be contrasted with those presented in Sec.
\ref{sec:bound} for bound material, where we analyse tracers in the core
of an HMNS produced by a BNS merger. Finally, in Sec.
\ref{sec:conclusions} we summarise our results and discuss their
applicability to future work. Unless otherwise specified, we use a system
of units such that $c = G = M_\odot = 1$, where $c$ is the speed of light
in vacuum, $G$ is the gravitational constant, and $M_\odot$ is the mass
of the Sun. We use Einstein's convention of summation over repeated
indices. Latin indices run over $1,2,3$, while Greek indices run over
$0,1,2,3$. The spacetime metric signature we adopt is $(-,+,+,+)$.

\section{Mathematical and numerical setup}
\label{sec:mans}

\subsection{Relativistic hydrodynamics and neutrino transport }

Einstein's theory of general relativity can be written as
\begin{equation}
R_{\mu\nu} - \frac{1}{2}R g_{\mu\nu} = 8\pi T_{\mu\nu}\,.
\end{equation}
In order to solve the above equations, we use the 3+1 decomposition 
\begin{equation}
ds^{2} = -(\alpha^{2} - \beta_{i}\beta^{i})dt^{2} + 2\beta_{i}dx^{i} dt +
\gamma_{ij}dx^{i}dx^{j},
\end{equation}
where $\alpha$ is the lapse, $\beta_{i}$ is the shift vector, and
$\gamma_{ij}$ is the 3-metric \cite{Rezzolla_book:2013}. We then
decompose the spacetime into a conformal-traceless 3+1 formulation known
as the \mbox{BSSNOK} formulation of the Einstein equations
\cite{Nakamura87, Shibata95, Baumgarte99}. This decomposition deals with
the spacetime, but to simulate BNS mergers, we require a description for
the underlying matter. To do so, we model the neutron stars as perfect
fluids, where the energy-momentum tensor is given by
\begin{equation}
T_{\mu\nu} := (e+p)~u_\mu u_\nu + p~g_{\mu\nu}\,,
\end{equation}
where $u^\mu:=dx^\mu/d\tau$ is the four velocity of the fluid with
position $x^{\mu}$ and proper time $\tau$, $e$ is the energy density, and
$p$ is the pressure. In order to include neutrinos, we must take into
account weak interactions, which can change the composition of the
material. Without weak interactions, the baryon number is conserved,
however when they are included, a source term must be added to the
conservation equation, which can be written as
\begin{equation}\label{eq:continuity}
  \nabla_\alpha (n_{e}\, u^\alpha) = R\,,
\end{equation}
where $n_{e}$ is the electron-number density and where $R$ is the net
lepton number emission/absorption rate per unit volume in the fluid
rest-frame. Likewise, the conservation of energy and momentum now
becomes, with the introduction of sources
\begin{equation}\label{eq:euler}
  \nabla_\beta T^{\alpha\beta} = Qu^{\alpha}\,,
\end{equation} 
and $Q$ is the net neutrino cooling/heating rate per unit volume in the
fluid rest-frame \cite{Radice2016}. A detailed discussion on the
computations of $Q$ and $R$ within the code employed here is contained in
Refs.~\cite{Galeazzi2013, Radice2016}. In order to close the system of
equations (\ref{eq:continuity}) and (\ref{eq:euler}) an equation of state
(EOS) of the form $p=p(\rho,\epsilon,Y_{e})$ is required, where $\rho$ is
the rest-mass density, $\epsilon$ is the specific internal energy, and
$Y_{e}$ is the electron fraction. The EOSs we use are the LS220 EOS of
\cite{Lattimer91} with a nuclear compressibility parameter $K =
220\ \mathrm{MeV}$ and the DD2 EOS \cite{Typel2010}. These EOSs have
been used extensively in numerous simulations, \eg \cite{Galeazzi2013,
  Hanauske2016, Radice2016, Baiotti2016} and provide useful test cases.
                                                                                
To solve the spacetime equations numerically we use the
\texttt{Mclachlan} code \cite{Brown:2008sb} which is part of the
\texttt{Einstein Toolkit} \cite{Loffler:2011ay}. The hydrodynamics are
solved using the \texttt{WhiskyTHC} code \cite{Radice2013b,Radice2013c}.
\texttt{WhiskyTHC} implements finite-volume and finite-difference with
high-resolution shock-capturing methods and has been used extensively to
study BNS mergers.

\subsection{Initial data and grid setup}

We have considered two types of initial data for the simulations
depending on whether our analysis focuses on unbound or bound material,
In both cases, however, the BNSs refer to an irrotational flow and were
computed using the multi-domain spectral-method code \texttt{LORENE}
\cite{Gourgoulhon-etal-2000:2ns-initial-data} under the assumption of a
conformally flat spacetime metric and employing the EOS at
beta-equilibrium. For the study of the \emph{unbound material} we used
quasi-circular initial data -- at least four orbits -- with an initial
gravitational mass of $M=1.35\,\Msun$ and initial separation of $45 \km$
with the LS220 EOS. For the bound material, on the other hand, we have
again chosen quasi-circular initial data with a gravitational mass of
$M=1.35 \Msun$ and initial separation of $45 \km$ but with the DD2 EOS.
The reason for choosing a different EOS is that DD2 is stiffer than LS220
and collapses to a black hole at a later time allowing sufficiently long
timescales to investigate the structure of the core before collapse.
Because the fluid that becomes unbound is located far away from the
merger site, \eg at distances $\gtrsim 300\km$ from the centre of the
grid, a higher resolution is desirable to better capture the fluid. For
this reason, a resolution of $\Delta h_5 = 0.15\,\Msun \approx 220\,{\rm
m}$ for the finest refinement level was used for the unbound material
evolution and $\pi$-symmetry imposed. For the bound material, resolution
is not as important for investigating some aspects of the fluid
properties, see, \eg \cite{Hanauske2016}, but for completeness $\Delta
h_5 = 0.15\,\Msun \approx 220\,{\rm m}$ was also used for the finest
refinement level but with no $\pi$-symmetry to ensure that the one-arm
stability is allowed to develop \cite{Radice2016a}. Both simulations had
five refinement levels, reflection symmetries, and an outer boundary of
$512\,\Msun \approx 760 \km$.

\subsection{Tracer evolution}

In order to follow the evolution of the fluid, we place tracers in the
fluid which are able to record the properties of the fluid at a given
point. These tracers are massless, see Sec~\ref{sec:tm}, and are
passively advected through the equation \cite{Foucart2014,Mewes2016}
\begin{equation}
\label{eq:tracer_advection}
\frac{d\vec{\boldsymbol{x}}}{dt} = \vec{\tilde{\boldsymbol{v}}}\,,
\end{equation}
where the velocity vector $\vec{\tilde{\boldsymbol{v}}}$ refers to the
three-velocity of the fluid with respect to the coordinates. In the 3+1
formalism this velocity is related to the fluid velocity through
\begin{equation}
\label{eq:tracer_evol}
\vec{\tilde{\boldsymbol{v}}} := \alpha\vec{\boldsymbol{v}}- 
\vec{\boldsymbol{\beta}}\,,
\end{equation}
with $\alpha$ the lapse, $\vec{\boldsymbol{v}}$ the three-velocity of the
fluid, and $\vec{\boldsymbol{\beta}}$ the shift vector. We solve Eq.
(\ref{eq:tracer_advection}) with a simple forward Euler scheme where the
$\Delta t$ is equal to the time-step of the finest grid. In our
simulation, this equation can be solved for a variable number of tracers
$N_{t}$ and we have chosen to set $N_{t}=10^5$ throughout. Furthermore,
since the tracers do not interact and are considered point particles they
can, in principle, occupy the same coordinates; in practice, however,
this does not happen in the simulations we have performed. As the tracers
pass through the grid, the value of various variables is recorded. Since
the variables are computed only on the grid points, and in general the
tracers will not be on the grid points, the desired properties are
interpolated to the $(x,y,z)$ position of the tracer. Specifically, the
most important quantities for nucleosynthesis are $(\rho, T, Y_{e})$,
which are evolved through the main evolution code.

\subsection{Tracer mass flux}\label{sec:tm}

As mentioned above, the main goal of studying the unbound material is to
use the tracers as input for nuclear-reaction networks. When the tracers
are post-processed through a nuclear-reaction network, the resulting
distribution of the abundancies must be weighted in some way. To do this,
a ``mass'' must be associated to each tracers. In contrast to SPH codes,
where the tracers explicitly represent the underlying fluid and can have
a mass, the tracers we employ are \emph{massless} and attention needs to
be paid when wanting to assign a ''mass`` to an individual tracer. In
some fully general-relativistic works, \eg Ref. \cite{Mewes2016}, tracers
have been assigned an associated mass; this concept, however, is
potentially misleading. First, since tracers play only a passive role
without a coupling to the fluid besides that of following the advection
equation (\ref{eq:tracer_advection}); assigning a mass to the tracers
breaks these assumptions. Second, if a tracer is initially associated
with a mass based on its surroundings, then at later times this mass
cannot represent the same mass it did initially since the underlying
properties will have changed. For example, if the initial mass is
calculated from the rest-mass density where the tracer is initially at,
then at a later time the density will have decreased significantly and no
longer represents the same mass as it did before. Finally, in order to
calculate a mass from a density, some volume must be specified and
tracers represent idealised point particles.

Having made these remarks, and since nucleosynthesis calculations require
masses to weight the different abundance curves, a scheme to assign a
``mass'' with the tracers is necessary. Our proposal for such a scheme
avoids all of the above issues by calculating the flux of the tracers
through a given 2-sphere of coordinate radius $R$, which we take to be
$200\,\Msun \approx 295 \km$ \cite{Radice2016,Bovard2017}. We can
therefore define the ``tracer mass current'' at a given iteration $i$ as 
\begin{equation}
\vec{\boldsymbol{j}}_{i} := 
\vec{\boldsymbol{j}}(\vec{\boldsymbol{x}},t_{i}) = \rho W (\alpha
\vec{\boldsymbol{v}}- \vec{\boldsymbol{\beta}}) =
\sum_{j}W_{j}(\vec{\boldsymbol{x}},t)\rho_{j}(\vec{\boldsymbol{x}},t)\, \vec{\boldsymbol{v}}_{j}
(\vec{\boldsymbol{x}},t)\,,
\label{eq:ji}
\end{equation}
where we sum over the tracers that cross through a surface of a given
radius during that iteration. Because we evaluate the flux far from the
BMP, the geometry is approximately flat, \ie $\alpha \simeq 1,
\vec{\boldsymbol{\beta}} \simeq 0$, so that the final
special-relativistic expression (\ref{eq:ji}) provides a good
approximation. As the information about the velocity, Lorentz factor, and
the rest-mass density are recorded, we have a way to associate a flux
with the tracers crossing the surface. Integrating over the surface at a
given time allows us to define a ``tracer mass flux" as
\begin{equation}
\dot{M}_{i}(t) = \oint \vec{\boldsymbol{j}}_{i}
\cdot \vec{\boldsymbol{n}}\sqrt{\gamma}\, dS =
\oint \vec{\boldsymbol{j}}_{i}
\cdot \vec{\boldsymbol{n}}\, dS\,,
\end{equation}
where we take the surface element on a sphere to be $dS=r^{2}\sin\theta d\theta
d\phi$ and again taking the special relativity expression since we are
sufficiently far away from the BMP. Then we are able to obtain a simple
differential equation
\begin{equation}
\label{eq:mass_flux}
\frac{dM_{\rm tr}}{dt} = \dot{M},
\end{equation}
for the tracer mass flux which can be integrated to obtain a mass
associated with the tracers. We stress that this ``mass'' is only valid
as the tracers pass through the surface and does not represent the true
mass unbound that would be measured when calculating the flux of the
underlying fluid through a sphere.  In Sec.~\ref{sec:ustats} we will
compare the mass flux computed via the tracers, \ie Eq.
(\ref{eq:mass_flux}), with the corresponding quantity computed using
standard hydrodynamical quantities, \eg the rest-mass current, and
demonstrate that this method can reasonably well approximate that of the
``exact'' answer from the underlying fluid. In contrast, for the bound
material, we are only interested in following the streamlines of the
fluid, which is the most fluid dynamical role that passively advected
tracers play. In this case, the mass of the tracers is irrelevant.

\section{Tracing unbound material}
\label{sec:unbound}

We start our discussion on the use of tracers by considering the case in
which they are employed to describe the dynamics of matter that is
gravitationally \emph{unbound}. Obviously, due to the finite
computational domain, tracers cannot escape further then the outer
boundary of $512\,\Msun$ and thus we must have a criterion that records
whether or not the tracers will escape to infinity. To do this, we follow
the geodesic criterion, \ie we consider a tracer to be unbound when the
corresponding fluid element has the covariant time component of the
four-velocity $u_{t}\le -1$, which corresponds to considering tracers as
moving on geodesics and reaching infinity with zero energy
\cite{Rezzolla_book:2013}. Other criteria could be used, such as the
Bernoulli criterion $h u_t \le -1$, which has been explored, for
instance, in Ref. \cite{Kastaun2014}. Here however, we have only made use
of the geodesic criterion as it is sufficiently robust, does not require
any tuning and generically does lead to rather conservative measurements
of the mass losses (a more detailed discussion will be presented in
Ref.~\cite{Bovard2017}).

In practice, every tracer records the value of the underlying fluid's
$u_{t}$ at that point and it is then determined in a post-processing
stage whether or not the tracer should be counted for comparison with the
fluid quantities. It is also important to ensure that the tracer
satisfies the geodesic criterion at all points, since it is possible for
the tracer to become unbound, then undergo shock heating and become
rebound. In a post-processing step, every tracer is checked to ensure
that once the tracer becomes unbound, it remains unbound. Finally, due to
the high velocity of the ejecta, tracers can easily reach the boundary of
the grid and, in this case, the tracer properties are set to atmosphere
values and dealt with in a post-processing step.

We next discuss an often neglected aspect of the treatment of tracers,
namely, their initial placement. Despite this being a very important
step, as it can significantly influence the overall results recorded by
the tracers, this is the first time that a systematic discussion is made,
at least to the best of our knowledge. The material that we present in
the following sections is admittedly detailed and the reader who is not
particularly interested in the more technical aspects can skip such
material and concentrate on the concluding remarks at the end of Sec.
\ref{sec:ustats} (see text in italics).

\subsection{Initial placement}

The final goal here is to accurately capture the essential properties of
the underlying fluid in the tracers. In order to properly represent the
underlying fluid, we need to ensure that we are sampling the fluid in a
sufficiently accurate manner. The important first step is to ensure that
when we initially place the tracers, we are sampling the material that is
most likely to become unbound. However, different placements of the
tracers initially can potentially lead to different representative
properties.  One potential placement scheme, which was followed in Refs.
\cite{Martin2015, Mewes2016}, is to place more tracers where there is
higher rest-mass density. This process is done randomly with more tracers
being placed at higher densities and fewer at lower densities. Foucart et
al. \cite{Foucart2014} have successfully used tracers to measure fluid
properties but no discussion is made on the initial placement procedure.
Finally, Wanajo et al. \cite{Wanajo2014} have used tracers in their study
of r-process nucleosynthesis, focussing however only on the $(x,y)$,
$(y,z)$, and $(x,z)$ planes; also in this case no discussion is made on
the impact of the placement of the tracers on their conclusions. Thus how
effective tracers are with respect to the underlying fluid and how their
results would change if they change the initial placement has been
unexplored. In what follows we discuss three different placement schemes
and the corresponding dynamics.

Before deciding \emph{where} to place the tracers, we must decide
\emph{when} they should be first distributed. Luckily, when studying BNS
merger, the most natural selection of placement time is straightforward:
at the merger. Prior to the merger, in fact, there is no mass outflow,
beyond some small spurious outflow due to initial conditions and the
inevitable and tenuous mass loss at the stellar surface. Another timing
option is to let the simulation evolve for a few milliseconds after
merger, which is when much of the dynamical ejecta is produced, and then
place the tracers to best capture this material. Once this optimal time
is fixed, we need a way to distribute the tracers. A priori, there is no
obvious ``best choice'' for what the correct initial placement of tracers
is. However, since we are also interested in using the tracer flux to get
a representative mass, choosing a scheme that is based on an initial
rest-mass density distribution does have merit. Furthermore, it is
reasonable to expect that low rest-mass density material around the
merging stars are going to be good candidates for material that will
become unbound. Conversely, we do not expect material at the high
densities in the core of the merger product to be ejected in contrast
with the schemes of \cite{Martin2015, Mewes2016}. Material inside the
neutron star that is between these two extremes can still be ejected due
to the complicated merging process, which can eject material from within
the neutron star.

In light of these considerations, at the chosen time during the
simulation we fix a Cartesian box surrounding the BMP and place tracers
at the grid points within this box, distributed according to a scheme
that is weighted with some probability distribution. In practice we have
considered four different options that we discuss in detail below.

\subsubsection{Tracers correlated/anticorrelated}

\begin{figure*}
\includegraphics[width=0.475\textwidth]{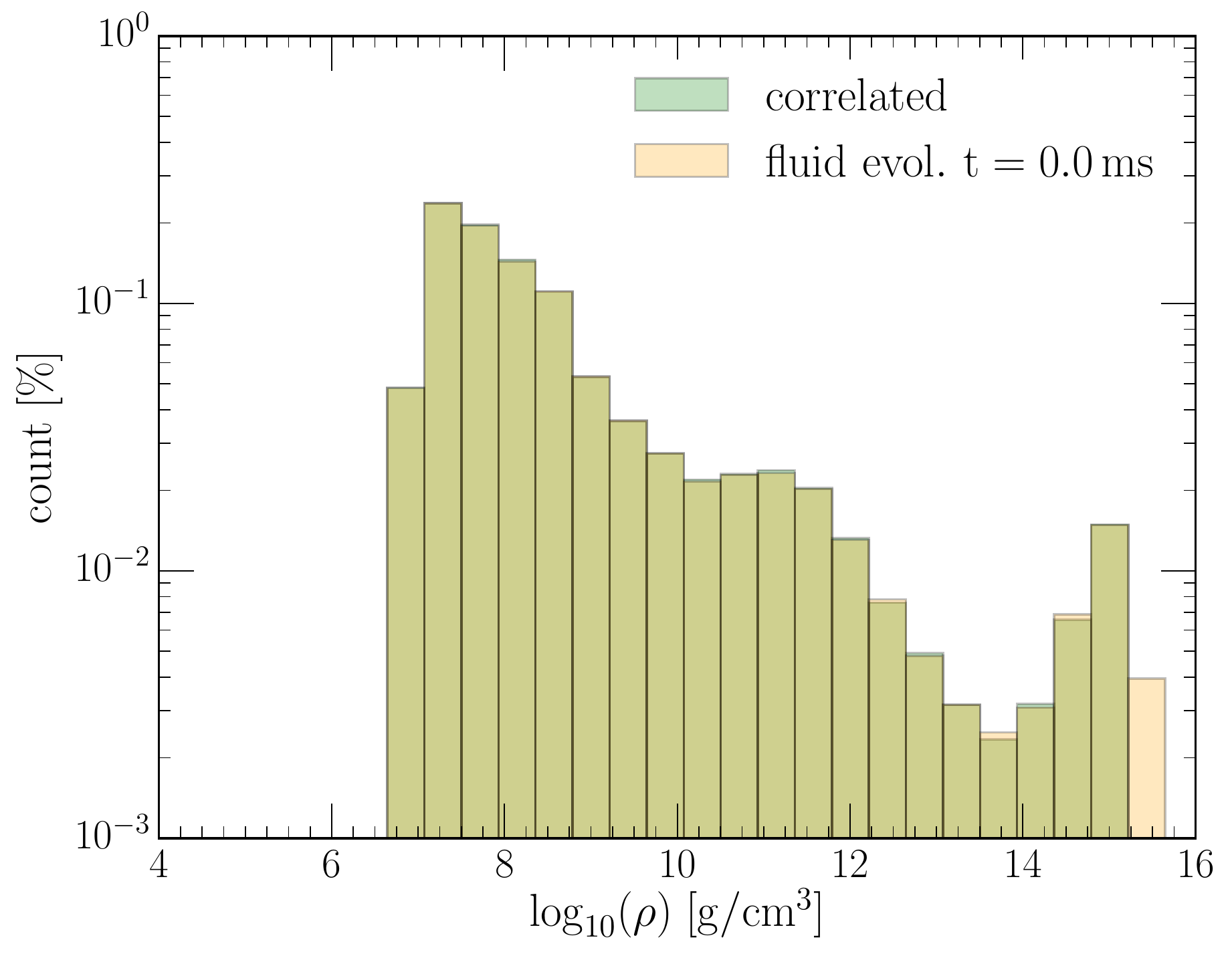}
\hskip 0.25cm
\includegraphics[width=0.475\textwidth]{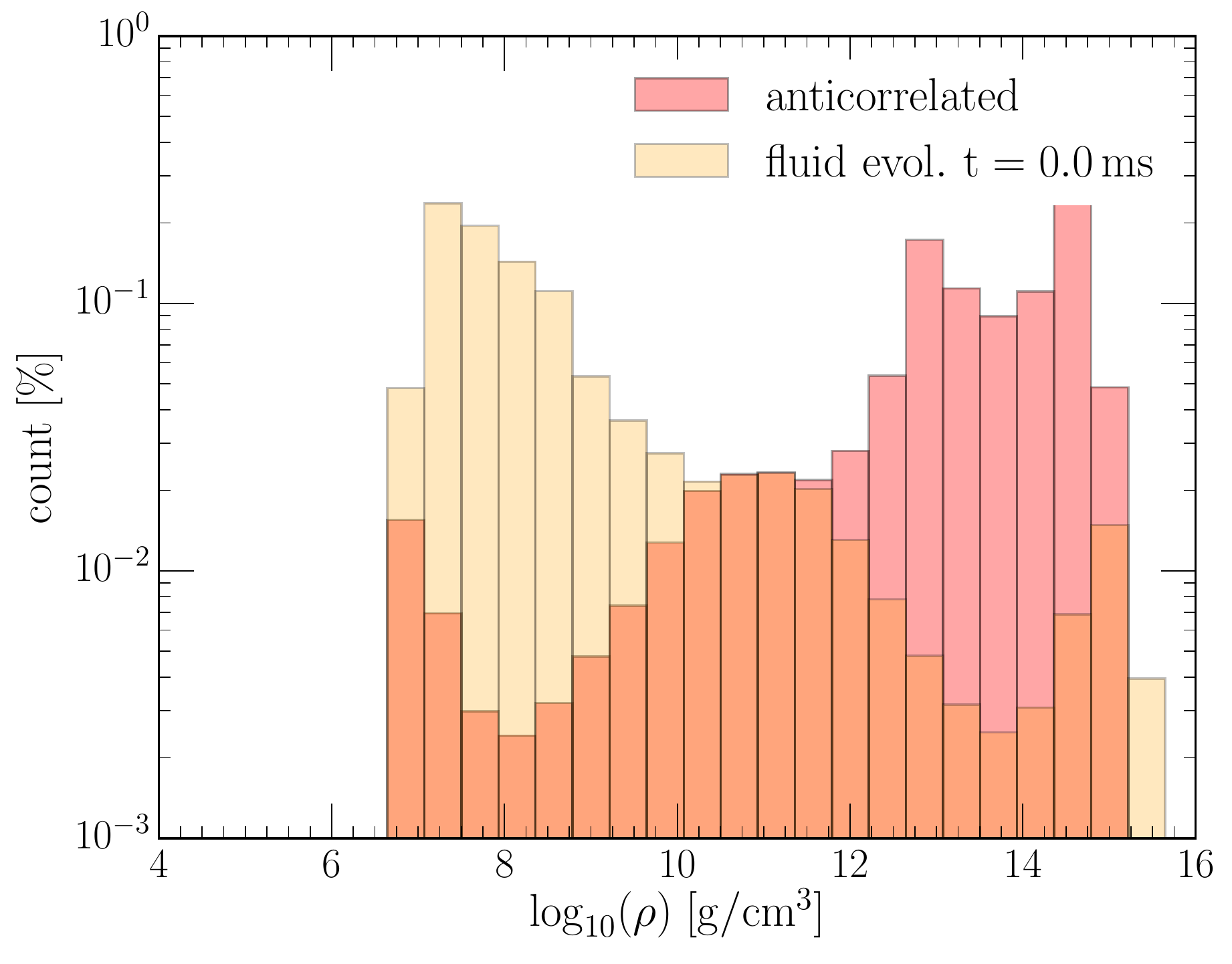}
\caption{Histograms showing the distribution of the underlying rest-mass
  density of the fluid (light-yellow shade) compared with the initial
  placement of the tracers for the ``correlated'' (green) and the
  ``anticorrelated'' (red). The histograms are normalised with respect to
  the total number of cells for the fluid and $10^5$ for the tracers. The
  placement is done at merger.}
  \label{fig:corranti}
\end{figure*}

Firstly, we consider the most natural probability distribution, namely,
placing tracers using a distribution function which is directly
correlated to the rest-mass density distribution at merger. In other
words, within our fixed box we count the number of cells within a given
rest-mass density range, then normalise by the total number of cells, and
finally bin them appropriately making sure that no more than one tracer
is assigned to a given cell.

In both panels of Fig.~\ref{fig:corranti}, the light-yellow shaded distribution
represents the underlying rest-mass density distribution produced by the fluid
evolution and calculated directly from the grid data. Note that the fluid
rest-mass density ranges from $\sim 10^{15} \gcm$ in the core out to $\sim
10^{7}\gcm$ in the material around BMP, which represents approximately $20\,\%$
of the rest-mass density cells. Note also that in our simulations the atmosphere
is set at $\sim 6\times 10^{3} \gcm$ \cite{Radice2013b, Radice2013c} and the
lowest rest-mass density of initial placement is at least three orders of
magnitude larger so no tracers are placed at the atmosphere. The left panel of
Fig.~\ref{fig:corranti} compares the rest-mass distribution computed from the
fluid evolution with the one adopted for the tracers, to which it is
\emph{``correlated''} within a rest-mass density range (light-green shade). More
specifically, the lower rest-mass density is set to the same as the fluid while
the upper rest-mass density is taken to be $10^{15}\gcm$. We then select $10^5$
cells randomly from the underlying fluid bins. The variations between the
``correlated'' distribution and that of the fluid are very small and are
negligible, with the only difference occurring just below $10^{14}\gcm$ where
the fewest number of fluid cells with that rest-mass density exist.

This behaviour should be contrasted with the one obtained when the tracer
distribution is \emph{``anticorrelated''} with the rest-mass density and
that is shown in the right panel of Fig.~\ref{fig:corranti} (light-red
shade). In practice, we first randomly selected from the higher densities
and then randomly filled in the lower densities, thus resulting in an
almost reflected distribution from the underlying distribution. This
procedure also saturates some of the higher rest-mass densities, for
which a tracer placed at every cell leads to a flattening of the
distribution in the high rest-mass density region. At lower densities,
on the other hand, we do see an increase in the distribution which is due
to the random selection.

\subsubsection{Tracers uniformly distributed}

\begin{figure*}
\includegraphics[width=0.475\textwidth]{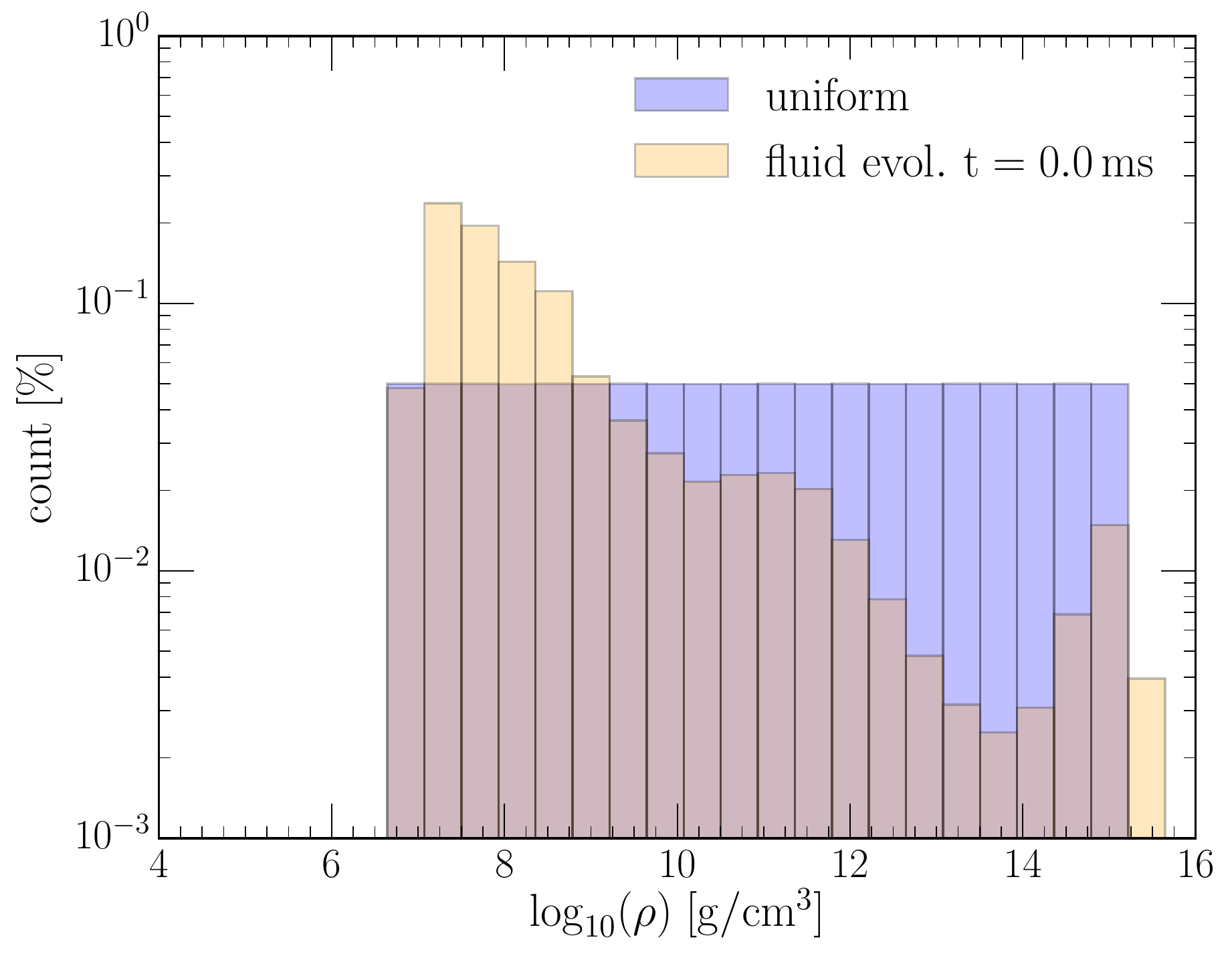}
\hskip 0.25cm
\includegraphics[width=0.475\textwidth]{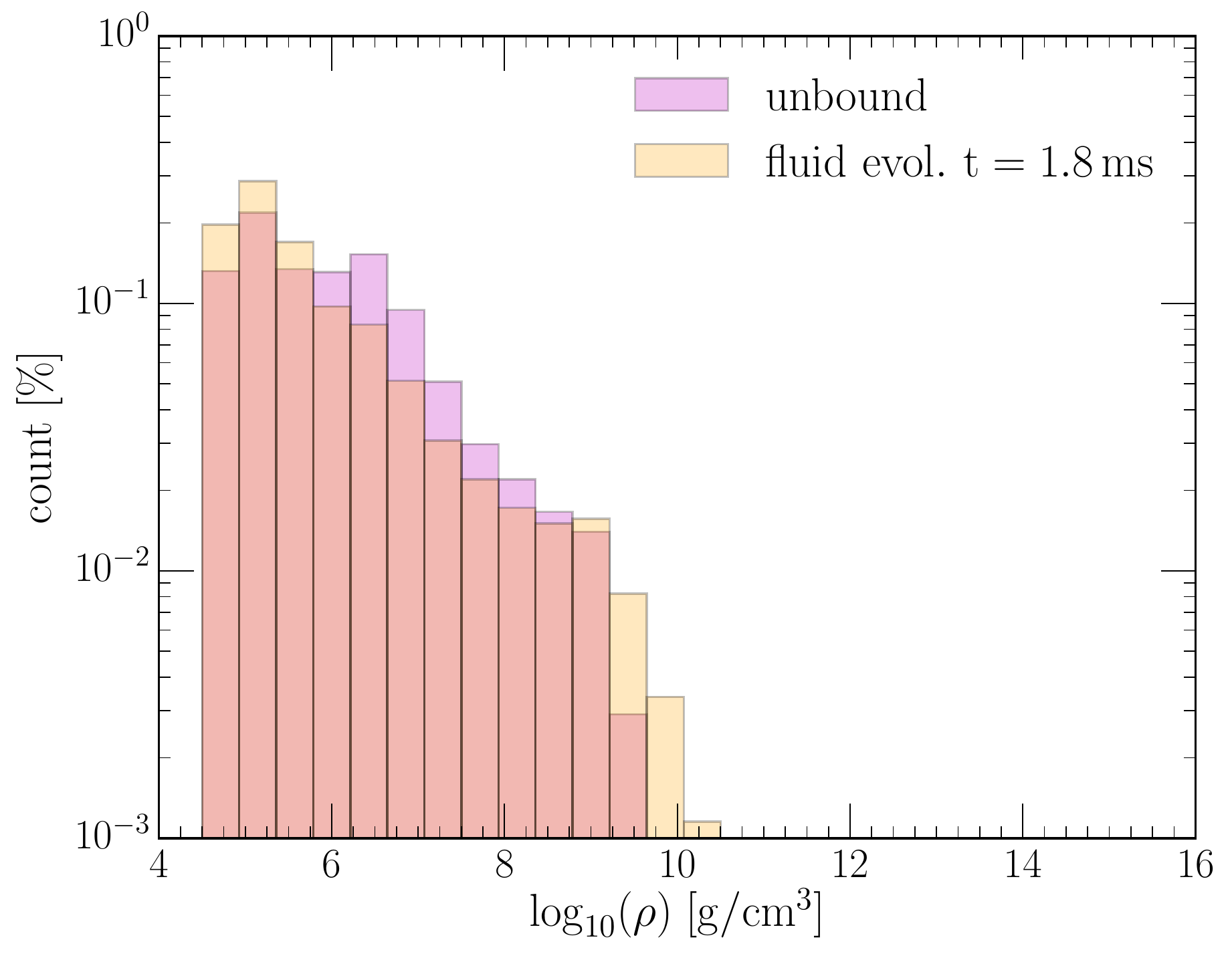}
\caption{\textit{Left:} Histograms showing the distribution of the
  underlying rest-mass density of the fluid (light-yellow shade) compared
  with the initial placement of the tracers for the uniform (blue). The
  histograms are normalised with respect to the total number of cells for
  the fluid and $10^5$ for the tracers. The placement is done at merger.
  \textit{Right:} The same as in the left panel but showing the
  distribution of the underlying rest-mass density of the fluid
  (light-yellow shade) compared with the placement of the tracers for the
  $u_{t}$ placement scheme (purple). The histograms are normalised with
  respect to the total number of cells for the fluid and $10^5$ for the
  tracers. The placement is made $1.8\mss$ after the merger.}
  \label{fig:inituni}
\end{figure*}

The third placement we consider at the merger is the uniform
distribution. In this case, we equally divide the tracers into the 20
bins that the underlying fluid was divided into. This results in a
distribution that does not follow the underlying rest-mass density
distribution at all, but instead places an equal number of tracers at all
densities. The resulting distribution is shown in the left panel of
Fig.~\ref{fig:inituni} (light-blue shade) and when comparing with the
underlying fluid (light-yellow shade) we can clearly observe that this
placement oversamples the higher densities (\ie $\gtrsim 10^{10} \gcm$)
while it undersamples the lower densities (\ie $\lesssim 10^{8} \gcm$).

\subsubsection{Tracers following unbound matter}

The last placement we considered is different from the previous three in
at least two important aspects. First, a proper choice needs to be made
about the time of placement. To this scope, we required that the initial
dynamical outflow had not yet passed the $200\,\Msun$ radius, which is
where we collect information about the underlying fluid through a
spherical surface placed to measure the outgoing flow
\cite{Radice2016}. This time worked out to be $1.8 \mss$ after the
merger; at this time the dynamical ejecta are close to, but have not
passed the $200 \Msun$ radius. Second, instead of placing the tracers
based on the rest-mass density, we placed them based on $u_{t}$ and only
considered cells where $u_{t}\le-1$, \ie at that time the cell
represented unbound material. Then, we randomly selected $10^5$ of those
cells and placed a tracer at that location. The resulting distribution is
shown in the right panel of Fig.~\ref{fig:inituni} (light-purple shade)
and it is interesting to note that even though we distributed the tracers
based on $u_{t}$, the distribution still has approximately the same
structure as that of the evolved fluid\footnote{Note that the
  distribution coming from the fluid evolution is different than that on
  the left panel of Fig.~\ref{fig:inituni} since it refers to a different
  time in the simulation.}. The main differences are that it overshoots
in the rest-mass density between $10^{6} - 10^{8} \gcm$ and does not
place tracers at higher densities of $10^{10}\gcm$. This is likely due to
the fact that higher rest-mass density material is closer to the BMP and
less likely to be unbound. Also note that the underlying fluid
distribution densities extend above rest-mass densities of $10^{10}\gcm$
but do not appear in the right panel of Fig.~\ref{fig:inituni} because
the latter maintains the same dynamic range of the other three panels in
Figs.~\ref{fig:corranti} and~\ref{fig:inituni}.

\subsection{Three-dimensional dynamics of unbound tracers}
\label{sec:3D_dyn}

\begin{figure*}
\includegraphics[width=1\textwidth]{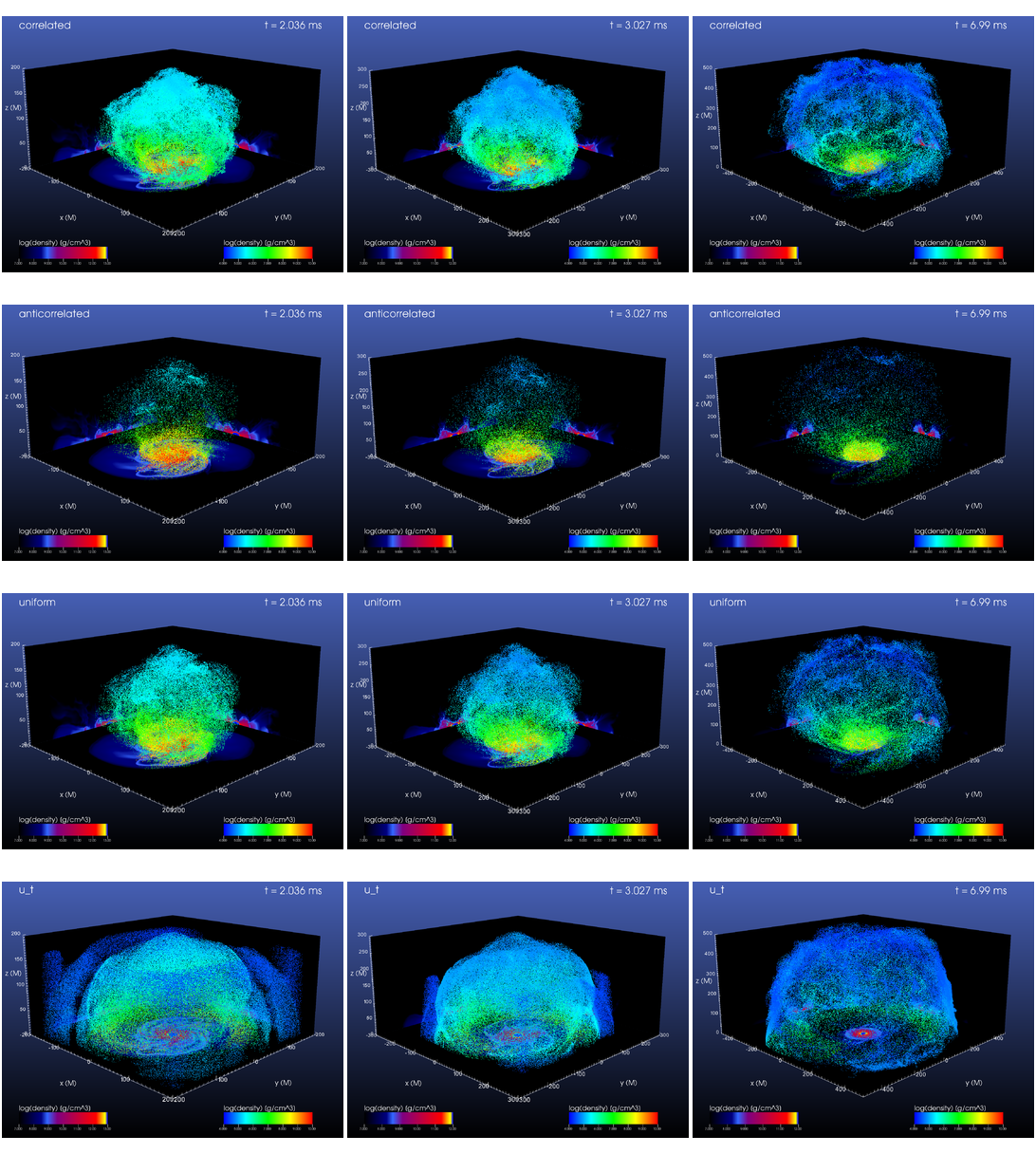}
\caption{Visualisation of the tracers for four different placement
  schemes (top to bottom) at three different times (left to right). From
  top to bottom: ``correlated'', ``anticorrelated'', ``uniform'', and
  ``unbound''. From left to right to $\sim 2,3,7\mss$ after merger. On
  the $(x,y)$, $(y,z)$, and $(x,z)$ planes are the shown the values of the
  rest-mass density using the colourbar in the left lower corner. A
  colourbar on the right lower corner is also used to visualise the
  rest-mass density of the fluid elements hosting the tracers. Note the
  different colour scales for the fluid and the tracers. }
  \label{fig:visit_dens}
\end{figure*}

\begin{figure*}
\includegraphics[width=\textwidth]{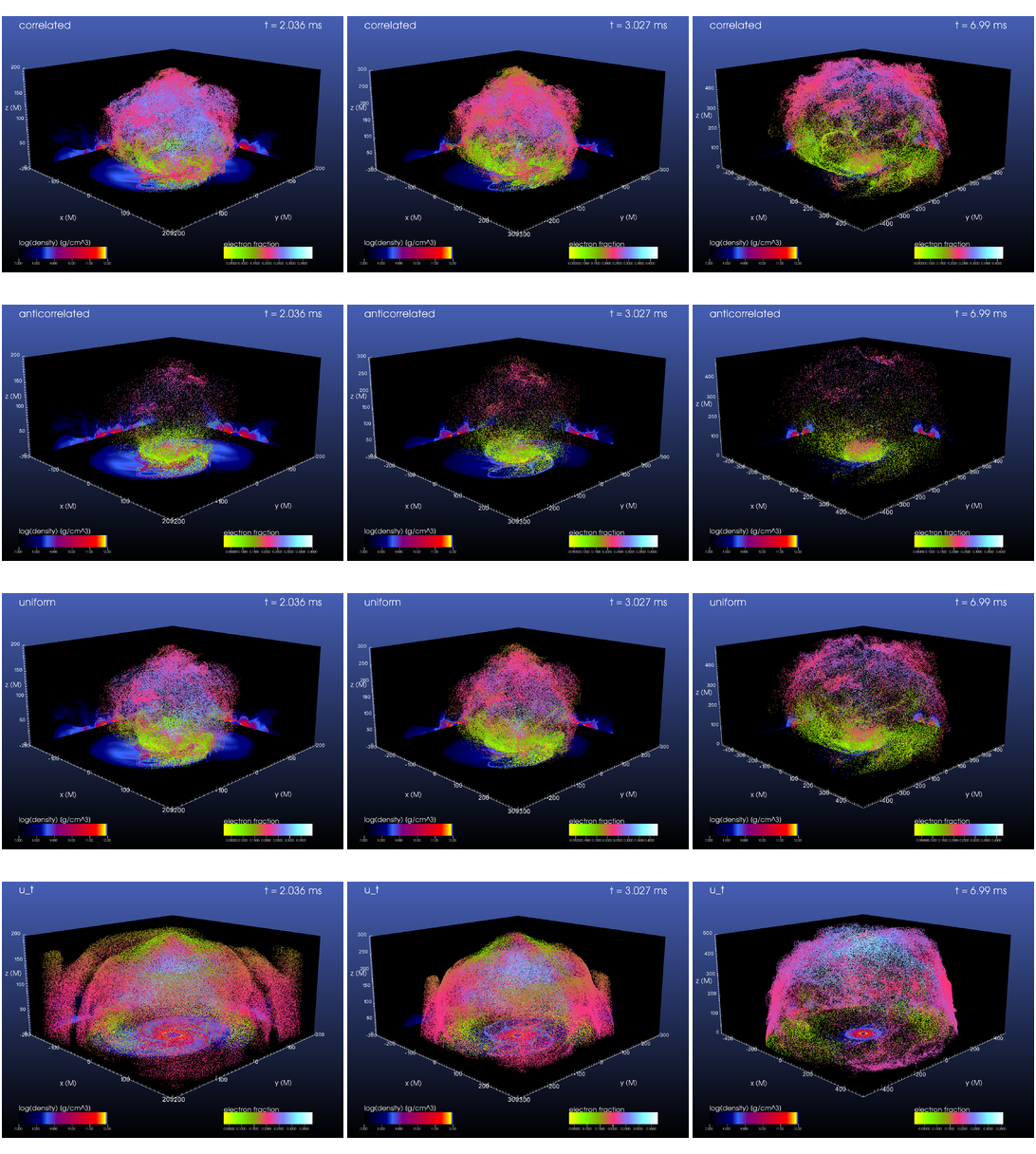}
\caption{Visualisation of the tracers for four different placement
  schemes (top to bottom) at three different times (left to right). From
  top to bottom: ``correlated'', ``anticorrelated'', ``uniform'', and
  ``unbound''. From left to right to $\sim 2,3,7\mss$ after merger. On
  the $(x,y)$, $(y,z)$, and $(x,z)$ planes are the shown the values of the
  rest-mass density using the colourbar in the left lower corner. A
  colourbar on the right lower corner is also used to visualise the
  electron fraction $Y_e$ of the fluid elements hosting the tracers.}
  \label{fig:visit_ye}
\end{figure*}

\begin{figure*}
\includegraphics[width=\textwidth]{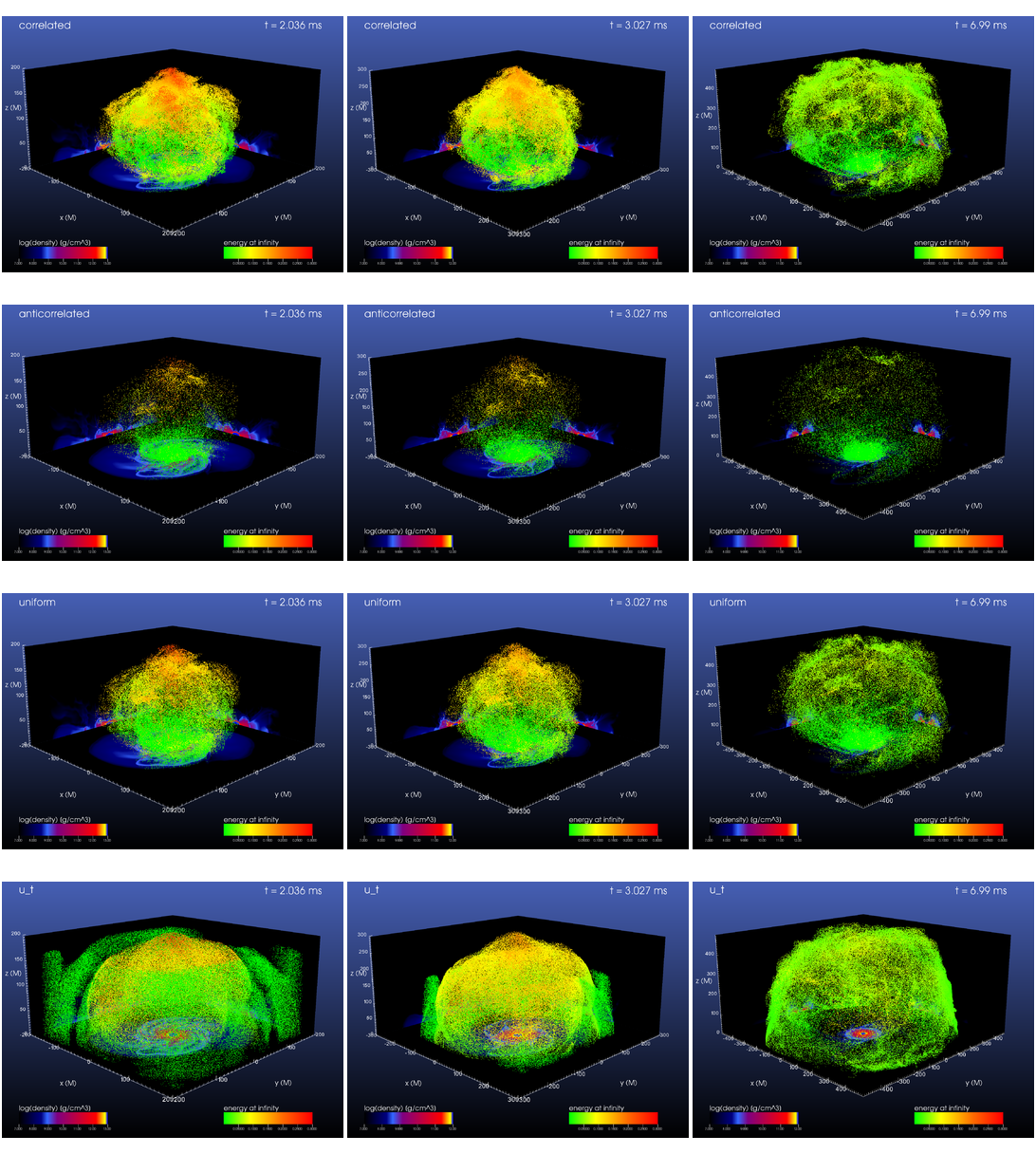}
\caption{Visualisation of the tracers for four different placement
  schemes (top to bottom) at three different times (left to right). From
  top to bottom: ``correlated'', ``anticorrelated'', ``uniform'', and
  ``unbound''. From left to right to $\sim 2,3,7\mss$ after merger. On
  the $(x,y)$, $(y,z$, and $(x,z)$ planes are the shown the values of the
  rest-mass density using the colourbar in the left lower corner. A
  colourbar on the right lower corner is also used to visualise the
  kinetic energy at infinity, $K:=-u_{t}-1$.}
  \label{fig:visit_unbd}
\end{figure*}

In order perform the nucleosynthesis calculations, the material that is
ejected has to have had sufficient time to become unbound and for this
reason all of the simulations presented here have been run for at least
$10\mss$ after merger. When using the geodesic criterion, in fact, there
is essentially no flux of unbound material or of tracers\footnote{This is
  not necessarily the case when using the Bernoulli criterion, which is
  less restrictive and allows for mass being ejected also at later times
  \cite{Bovard2017}.} through a 2-sphere of radius $200\,\Msun$; $10\mss$
are also a sufficient time for the tracers and the fluid to reach the
outer boundary of our computational domain. In Fig.~\ref{fig:visit_dens}
we show a visualisation of the tracers for the four different placement
schemes discussed above (from top to bottom: ``correlated'',
``anticorrelated'', ``uniform'', and ``unbound'') at three different
times corresponding roughly to when the unbound material passes through
spheres of radii $200,\, 300,\, 500\,\Msun$ (\ie $t \sim 2, 3,$ and
$7\mss$ after the merger, respectively). On the $(x,y)$, $(y,z)$, and
$(x,z)$ planes are the shown the values of the rest-mass density using
the colourbar in the left lower corner. A colourbar on the right lower
corner is also used to visualise the rest-mass density of the fluid
elements hosting the tracers. Note the different colour scales for the
fluid and the tracers.

For the case of ``correlated'' tracers (first row), we can see the presence of a
large number of tracers at high latitudes, which correspond to those initial
tracers that are at low rest-mass density. This is to be contrasted with the
dynamics of ``uncorrelated'' tracers (second row), which shows the
anticorrelated placement which has significantly fewer tracers at higher
latitudes and closer to the BMP, simply as a result of having downsampled fluid
elements at high rest-mass densities and oversampled those at high rest-mass
density. Not surprisingly, the ``uniform'' tracers strikes a place between the
two (third row). Finally, the case of ``unbound'' tracers (fourth row) shows
a very different structure, where it is possible to clearly distinguish the
different ``waves'' of matter ejected dynamically (left and central columns) and
where an almost spherically ``dome'' develops over the BMP product and which is
not present in the other placements. Note that this should not be interpreted as
an indication that the mass outflow is spherically symmetric as the tracers in
this case are not a faithful description of the rest-mass density. Indeed, when
using the colourbar to track the actual rest-mass density, it is possible to note
that high rest-mass density material (coloured red) is ejected mostly near the
equatorial plane, while low rest-mass density material (coloured blue) is
ejected at higher latitudes. Hence, the ejected material has a strong angular
dependence. This is particularly important in view of the previous studies of
nucleosynthesis with tracers carried out in Ref. \cite{Wanajo2014}, where the
tracers were analysed only on the $(x,y)$, $(y,z)$, and $(x,z)$ planes, and of
the fact that the angular dependence of the tracers can change the potential
observed properties (see Refs. \cite{Martin2015, Martin2015a} for a discussion).
The snapshots in Fig.~\ref{fig:visit_dens} demonstrate that there is significant
material that lies outside these planes; indeed, as we will further discuss in
Sec.~\ref{sec:ustats}, about $50\,\%$ of the mass ejected lies within 10 degrees
of the orbital plane, while the remaining $50\,\%$ occurs at higher angles from
the equator.

Figure~\ref{fig:visit_ye} is similar to Fig.~\ref{fig:visit_dens} but we
use the colourbar to represent the value of the electron fraction $Y_{e}$
carried by the various tracers. In all placements, we see a similar
distribution of the electron fraction, with low $Y_{e}$ values near the
equator and increasingly high values near the poles. For all
distributions, this increase in $Y_{e}$ is correlated with a decrease in
the rest-mass density. Thus despite having potentially high-$Y_{e}$
ejecta in the polar regions, it is associated with only a modest amount
of ejected matter. Overall, the snapshots indicate that there will be a
strong angular dependence of the r-process nucleosynthesis; a detailed
discussion of this will be presented in an accompanying paper
\cite{Bovard2017}. Finally, Fig.~\ref{fig:visit_unbd} is similar to the
two previous ones but refers now to the kinetic energy at infinity, \ie
$K := -u_{t}-1$, so that $K \ge 0$ corresponds to $u_{t} \le -1$.
Clearly, all snapshots show that, independently of the placement
criterion chosen, the material near the equator is just unbound, but, as
the latitude increases, the material becomes more unbound, increasing to
a maximum at the poles. Similarly, at late times, we can see that all the
high-energy material has already been ejected in the violent dynamics
accompanying the early postmerger and that and most values of $K$ are
close to zero. In summary, when analysed in a combined manner the
three-dimensional dynamics of the unbound tracers shown in
Figs.~\ref{fig:visit_dens}--\ref{fig:visit_unbd} reveals that the ejected
matter near the poles has lower densities, but larger values of electron
fraction and kinetic energy, and such a picture has implications for the
detection and observation of kilonova \cite{Bovard2017}.

\subsection{Distribution dynamics of unbound tracers}
\label{sec:spacetime_dyn}

\begin{figure*}
\includegraphics[width=0.5\textwidth]{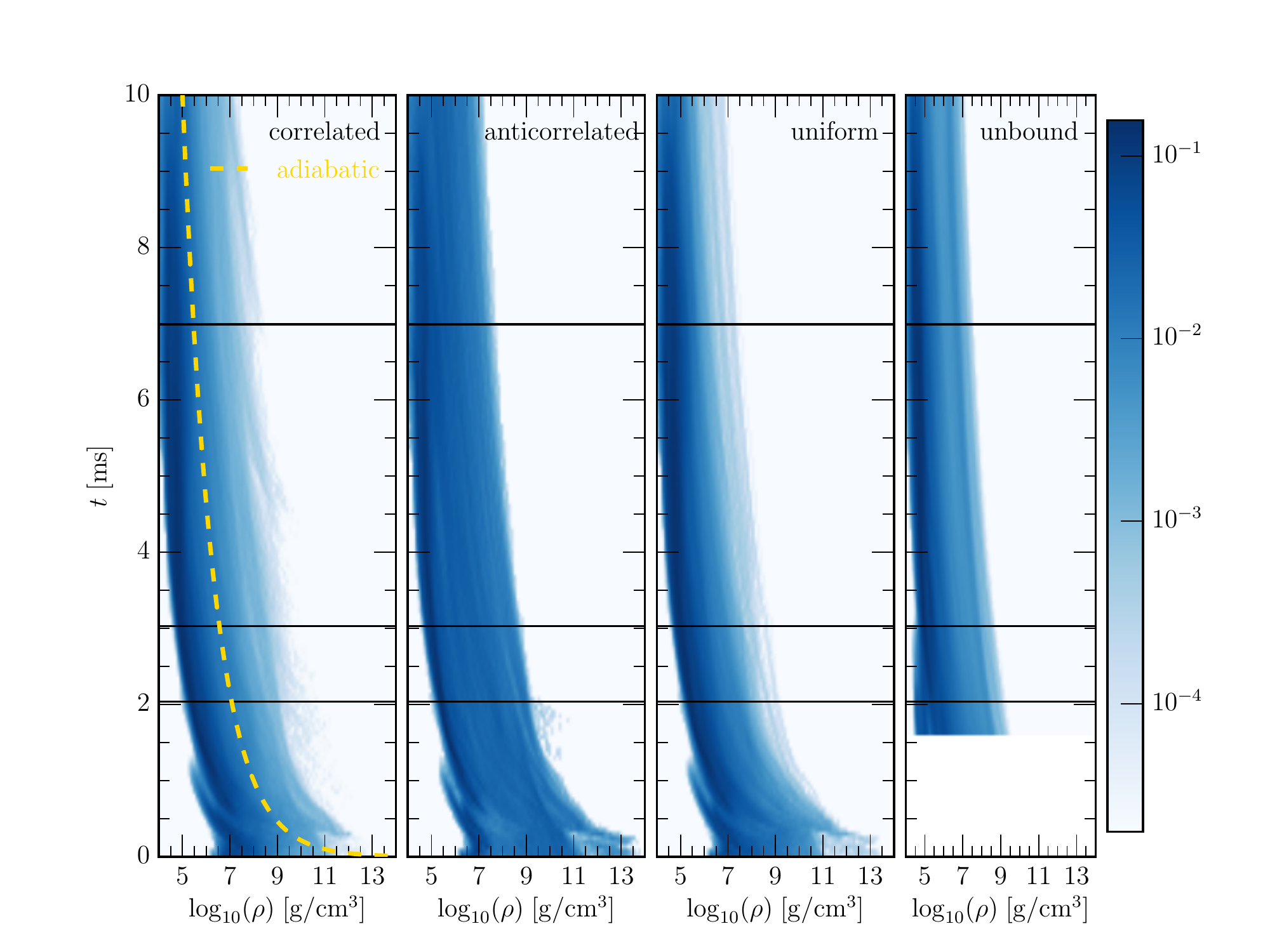}
\includegraphics[width=0.5\textwidth]{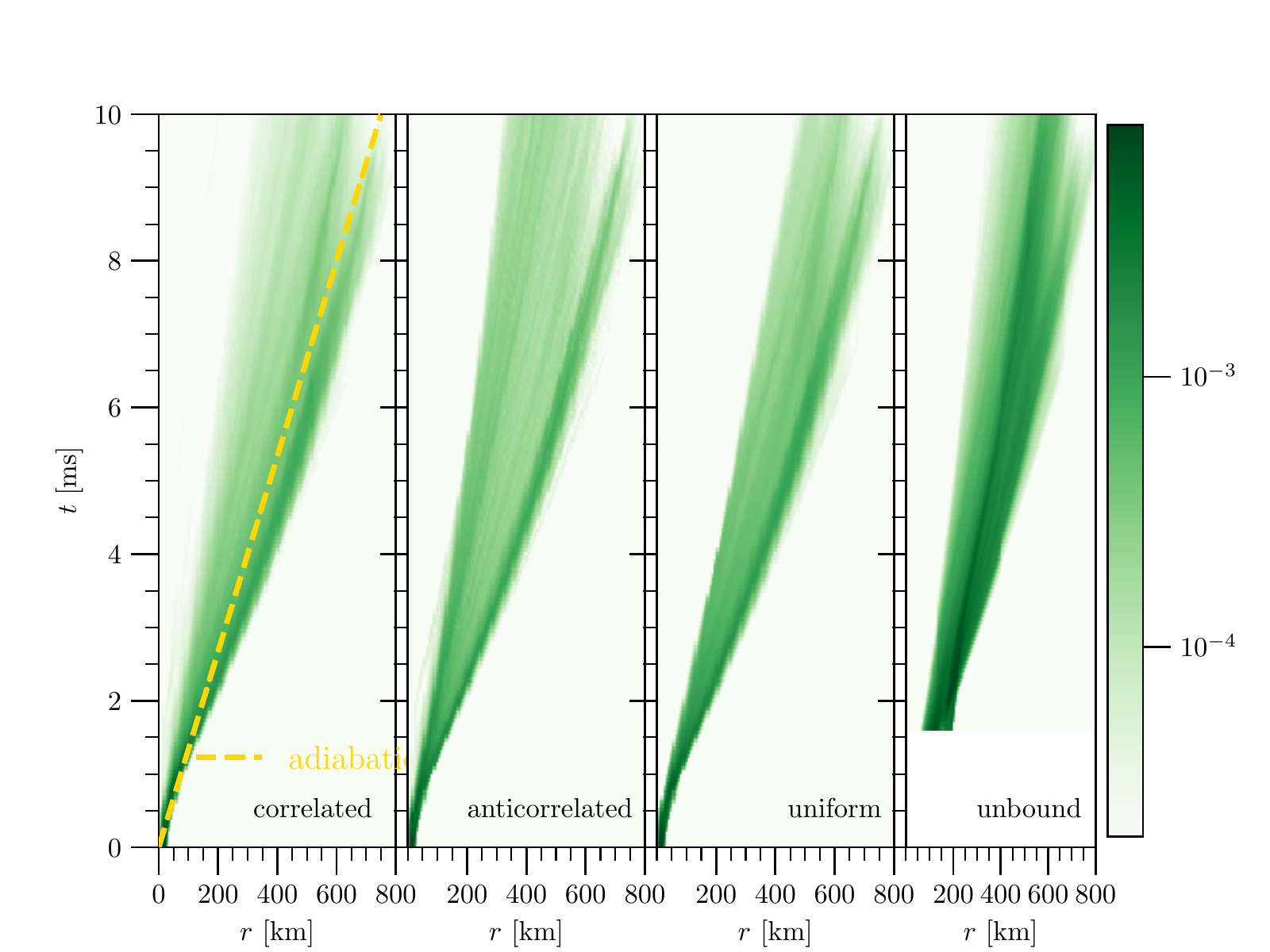}
\center
\includegraphics[width=0.5\textwidth]{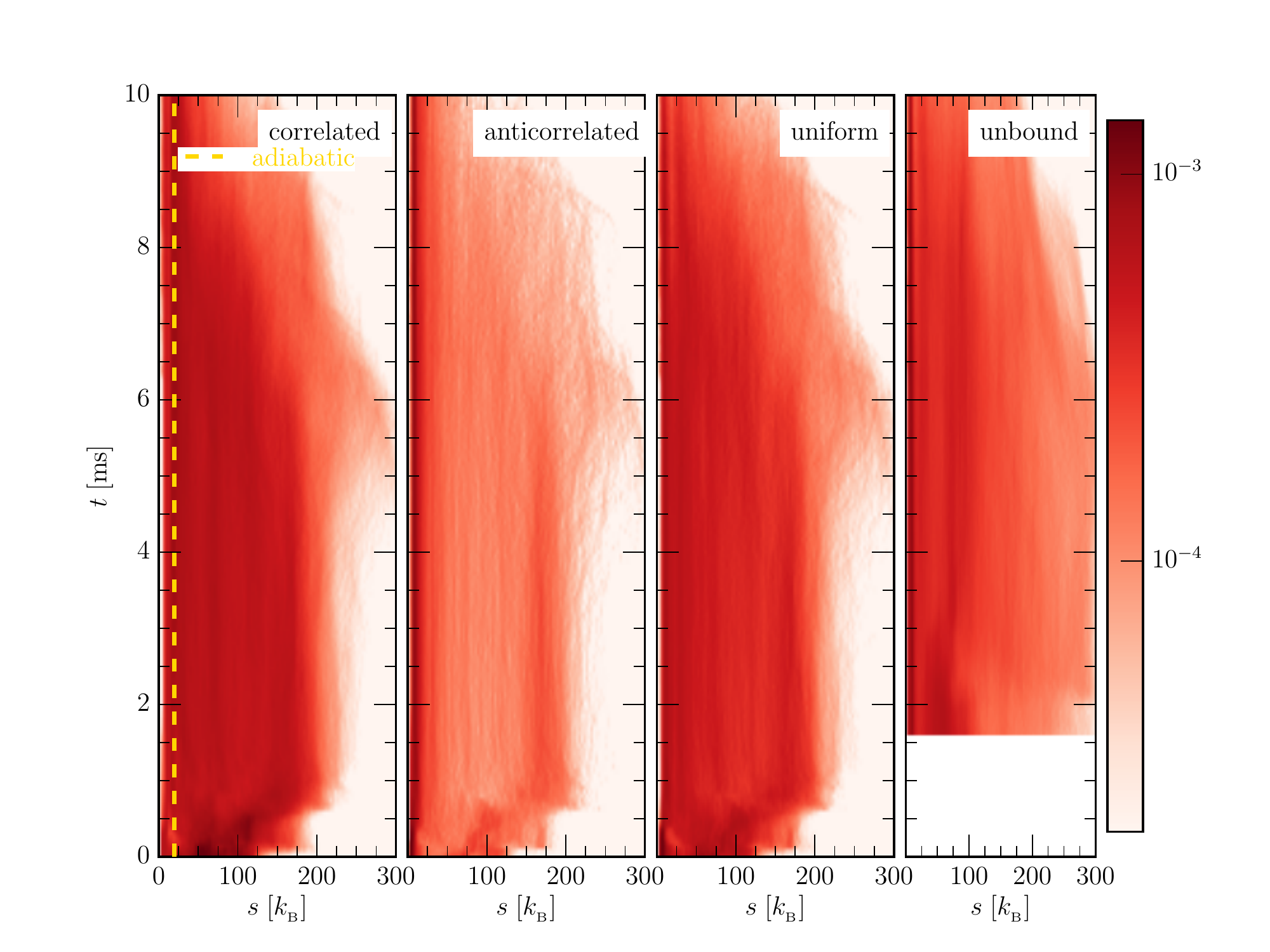}
\caption{Evolution of the distribution functions for three representative
  quantities: the rest-mass density, the radial position and the specific
  entropy. From left to right are the ``correlated'', ``anticorrelated'',
  ``uniform'', and ``unbound'' placements. The yellow lines represent the
  adiabatic expansion for each of the plotted quantities. Note that the
  ``unbound'' placement starts $1.8\mss$ after the merger. Horizontal
  black lines correspond to the times shown in
  Figs.~\ref{fig:visit_dens}--\ref{fig:visit_unbd}.}
  \label{fig:spacetime}
\end{figure*} 

A key assumption often made in studying the outflow material in r-process
nucleosynthesis and in kilonova modelling \cite{Li1998, Grossman2014}, is
that the unbound material is expanding adiabatically, \ie that the
internal energy remains constant. In addition, the material is assumed to
expand radially, so that the rest-mass density decreases in time as $\sim
t^{-3}$. While all reasonable, our use of tracers allows us now to test
these assumptions. We do this by reporting in Fig.~\ref{fig:spacetime}
the time evolution of the distribution functions of representative
quantities, \ie the rest-mass density, the radial position, the specific
entropy\footnote{A diagram showing the evolution of the distribution
function of the radial position of the various tracers effectively
represents a spacetime diagram, thus offering the opportunity to
visualise the (radial) worldlines of the tracers.}. In these plots, the
worldlines of the various tracers are marked with small dots whose colour
marks the corresponding fraction; as a result, a dark/light trajectory
will indicate that a large/small number of tracers with the corresponding
quantity having the reported values at a given time. 

In the top left panel of Fig.~\ref{fig:spacetime} we show the evolution
diagram of the evolution of rest-mass density, with the dashed-gold curve
representing adiabatic expansion, $\rho \sim t^{-3}$. The four sub-panels
distinguishing the various criteria adopted for the initial placement,
while the horizontal black lines correspond to the times shown in
Figs.~\ref{fig:visit_dens}--\ref{fig:visit_unbd}. In all cases, the
general trend of the adiabatic expansion can be seen very clearly and
involves both tracers with high rest-mass density and tracers with low
rest-mass density. However, it is important to note that while the
general trend is for an adiabatic expansion, not all tracers follow this
behaviour. For example, in the case of ``correlated'' tracers, the panel
shows that there is a small fraction of tracers (light blue)
corresponding to higher rest-mass densities that does not always follow
the adiabatic expansion prior. Another feature is the number of tracers
that reach close to atmospheric values and that eventually represents a
large fraction of the total number (dark blue). In all placements
considered there is a clear gradient of colours with increasing rest-mass
density. An exception to this behaviour is offered by the case of
``anticorrelated'' tracers, where the fraction gradient is not as
strong. This is a manifestation of the initial placement scheme: there
are more tracers at high rest-mass densities and of those that become
unbound, they represent a significant fraction of the unbound material.

Similarly, in the top right panel of Fig.~\ref{fig:spacetime} we show a
standard spacetime diagram of the radial distance from the origin of the
tracers; note that the ``unbound'' placement starts $1.8\mss$ after the
merger. In the adiabatic approximation, the expansion radius $r$ should
scale linearly with time, \ie $r\sim t$, and the figure shows clearly
that most tracers do follow a very close to radial ejection. In addition,
we can see the wide distribution of the tracers. Our domain is
approximately $750 \km$ wide from the origin in each direction and the
tracers have a radial distance ranging from $300 \km$ up to $800 \km$ at
any given time. This spatial distribution has been discussed in terms of
the neutrino-drive wind ejecta \cite{Martin2015a} but has not been
explored in the dynamic ejecta case in general relativity.

Finally, in the bottom panel of Fig.~\ref{fig:spacetime} we show the
evolution of the distribution function for the specific entropy, such
that an adiabatic expansion would simply be represented by a vertical line
for each tracer. The panel allows us to clearly note that indeed many of
the tracers obey this adiabatic expansion, but not all. For example, it
is clear that at high entropies the tracers do not expand adiabatically
and that tracers can actually increase their entropy, for instance as a
result of additional shocks with the outgoing matter. Hence, this panel
illustrates that the assumptions made that the ejected matter follows an
adiabatic flow is a very valid approximation for much of the unbound
material, but in order to have the most robust input data for the nuclear
network, one needs to have the full history of the fluid element. This is
in contrast with the analysis of \cite{Radice2016}, where the unbound
material was expanded adiabatically from the surface of the outflow
sphere.


\begin{figure}
\includegraphics[width=0.5\columnwidth]{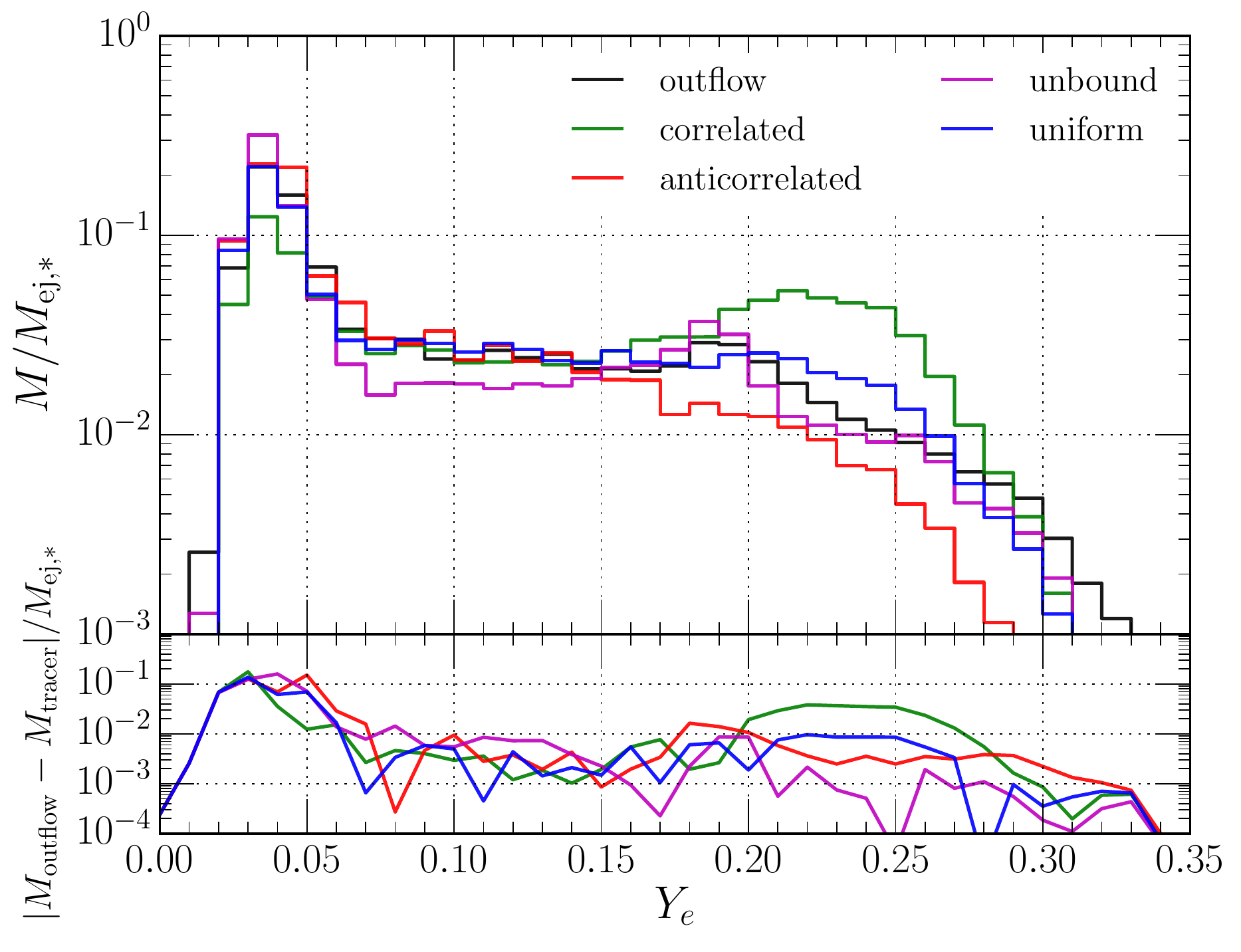}
\includegraphics[width=0.5\columnwidth]{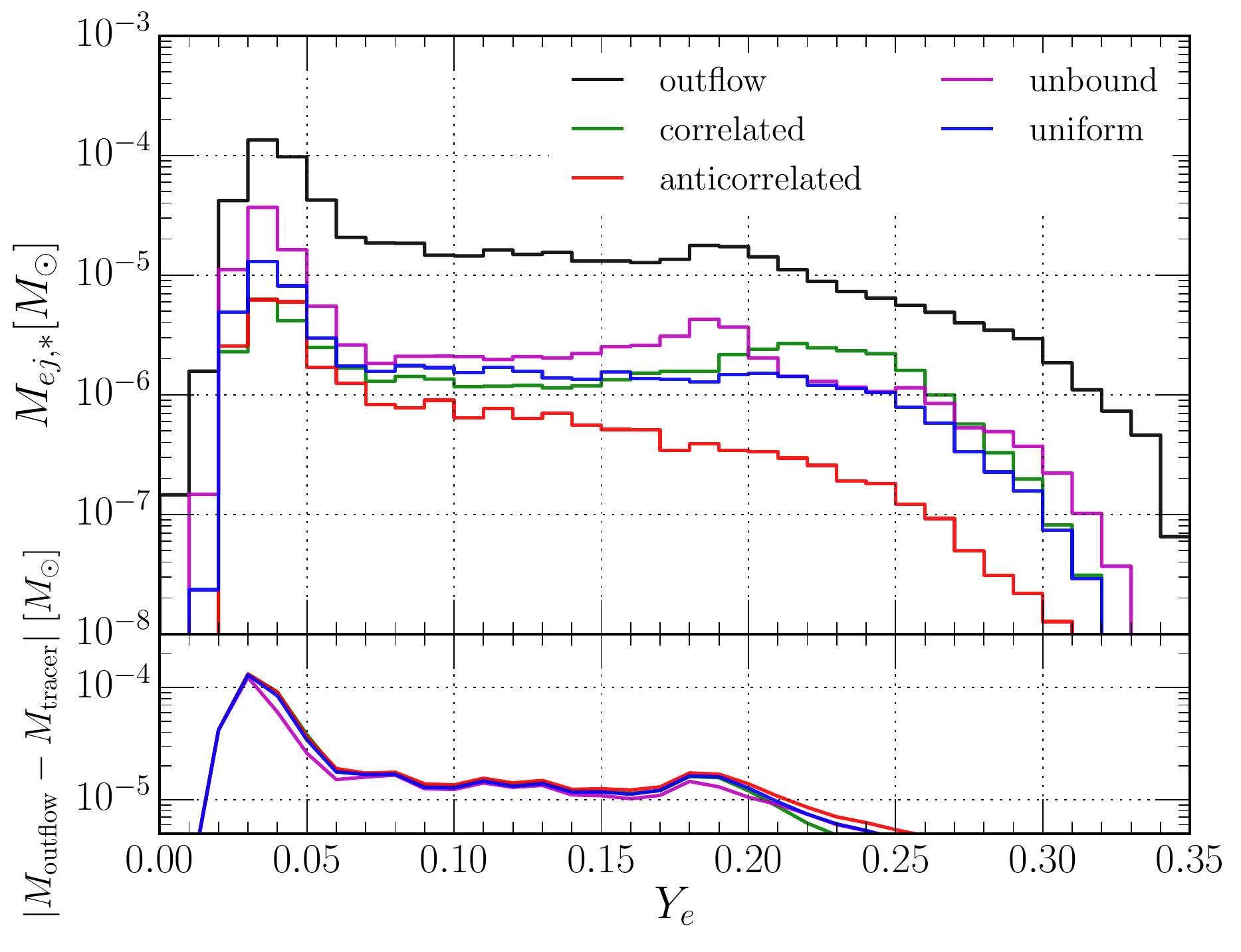}
\caption{\textit{Left panel:} Comparison between the distributions in
electron fraction as computed with the tracer mass-fluxes through the
surface at $200\,\Msun$ as computed with the different placement schemes
(coloured lines) and the outflow (black line). Because the ejected masses
are different in the various cases, all curves have been normalised to
the masses of their respective total values. Shown in the bottom part is
the relative error, which is smaller for the uniform distribution
placement. \textit{Right panel:} The same as in the left but when the
distributions are absolute and not normalised.} \label{fig:comp_methods}
\end{figure}

\subsection{Tracer and fluid information} \label{sec:ustats}

In the previous sections we have illustrated how different choices for
the placement of the tracers can lead to different dynamics of the
tracers and hence to different measurable quantities. We now address the
issue of which of these possible choices should be the recommended one in
numerical simulations in order to most accurately represent the
underlying fluid. In particular, to determine the efficacy of the
different placements, we have compared the tracer distributions of the
most important thermodynamical and physical quantities, \ie electron
fraction, entropy, energy at infinity, angular distribution of the mass
flux, with the corresponding distributions from the underlying fluid
evolution; hereafter we refer to the latter as the (fluid) ``outflow''.
As in the previous discussions, we compute our distributions in terms of
the fluxes across a spherical detector at $200 \Msun$ from the origin and
consider only material that is unbound according to the geodesic
criterion.

Perhaps the most important quantity of all thermodynamical quantities is
the electron fraction $Y_{e}$, which measures how neutron rich the
material is. The value of the electron fraction relates to which
r-process elements are created and what kind of kilonova is produced.
For these reasons, it is the quantity which we will use to determine
which placement best matches that of the underlying fluid. In
Fig.~\ref{fig:comp_methods} we plot the distributions of the electron
fraction using the different placement prescriptions (marked with lines
of different colour) along with the distribution obtained with the fluid
``outflow'' (black). In the bottom panel we also plot the error between
each of the tracer prescriptions with that of the ``outflow''. In the
left panel, in particular, all curves are normalised to the amount of
rest-mass ejected in the various cases, $M_{\rm ej,*}$, so that the
integral of the various distributions is unity. In the right panel of
Fig.~\ref{fig:comp_methods} such a normalisation is not used and the
distributions refer to the absolute amount of ejected matter for the
different distributions.

Overall, the fluid-outflow distribution (black solid line) shows that the
electron fraction has a main peak at $Y_{e} \sim 0.04$ and a secondary
one at $Y_{e} \sim 0.19$. Concentrating on the left panel, we clearly
note that the ``anticorrelated'' and ``uniform'' placements both give the
correct value for the $Y_{e}\sim 0.04$ peak with a relative error of
$\sim 10^{-2}$ and $10^{-3}$ respectively. On the other hand, the
``unbound'' ($u_{t}$) and the ``correlated'' placements over- and
underestimate it respectively by roughly the same amount of $10^{-1}$. In
particular, for the secondary peak, the ``correlated'' placement
overestimates the peak significantly with an error of about $50\%$; a
much better agreement is obtained with the ``unbound'' and ``uniform''
distributions, with errors that are of a few percent only. In the
intermediate electron-fraction range, \ie for $0.04 \lesssim
Y_{e}\lesssim 0.19$, all placements provide an accurate description, with
the only exception of the ``unbound'' placement, which tends to
underestimate the underlying fluid. Hence, all things considered, we can
conclude that the initial tracer placement that provides the best match
with the consistent fluid evolution is the \emph{``uniform''}
distribution.

\begin{figure*}
\includegraphics[width=0.5\columnwidth]{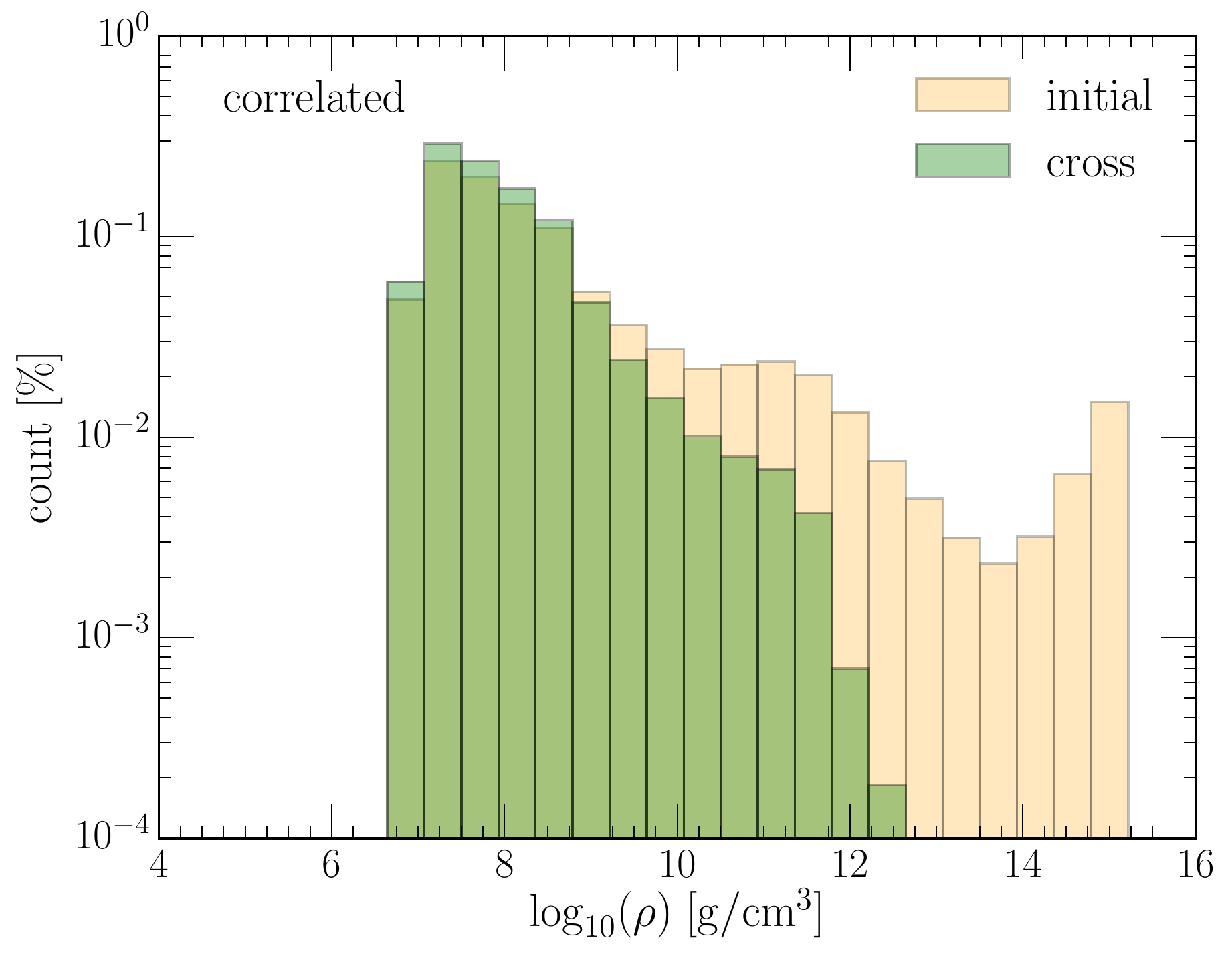}
\includegraphics[width=0.5\columnwidth]{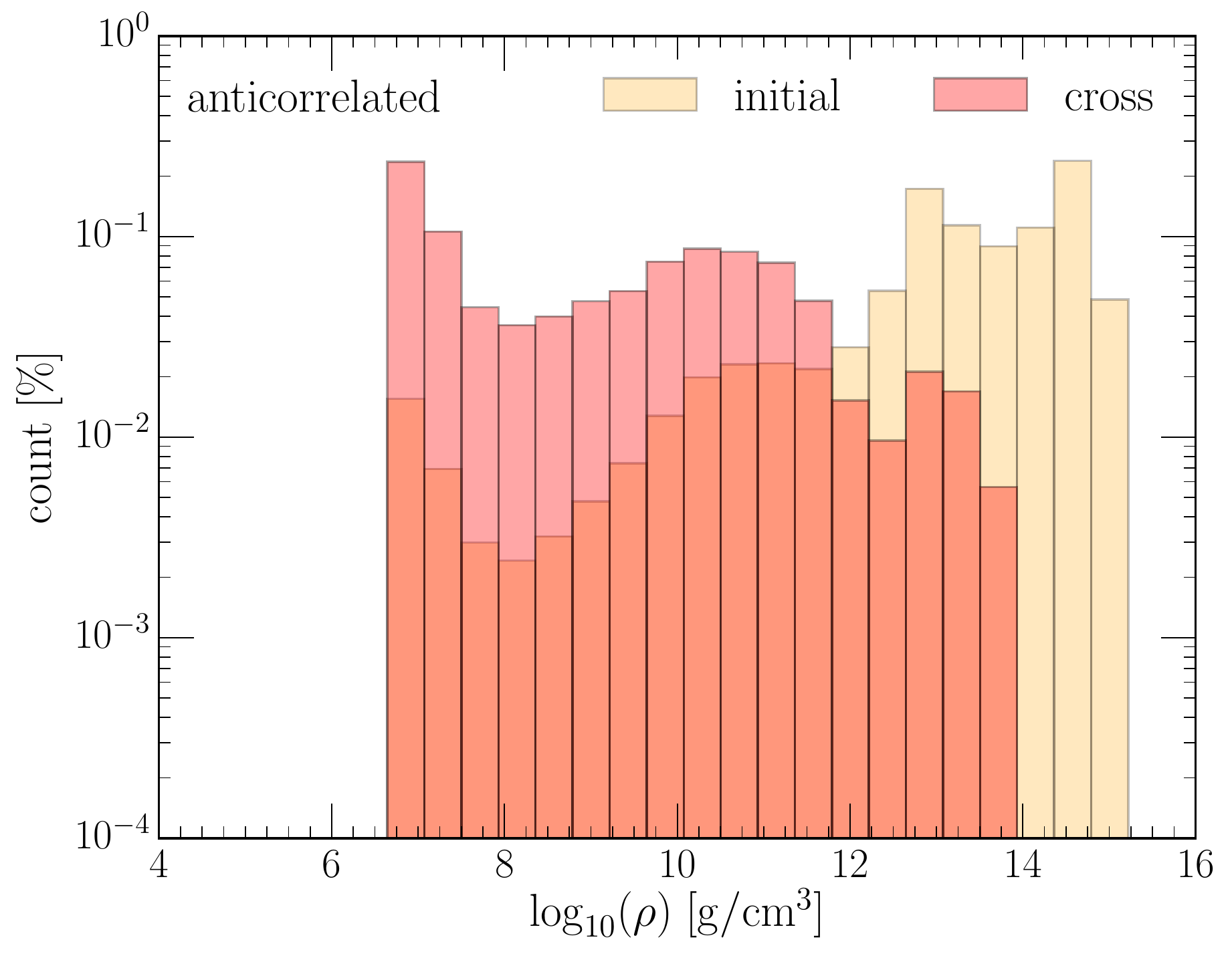}
\includegraphics[width=0.5\columnwidth]{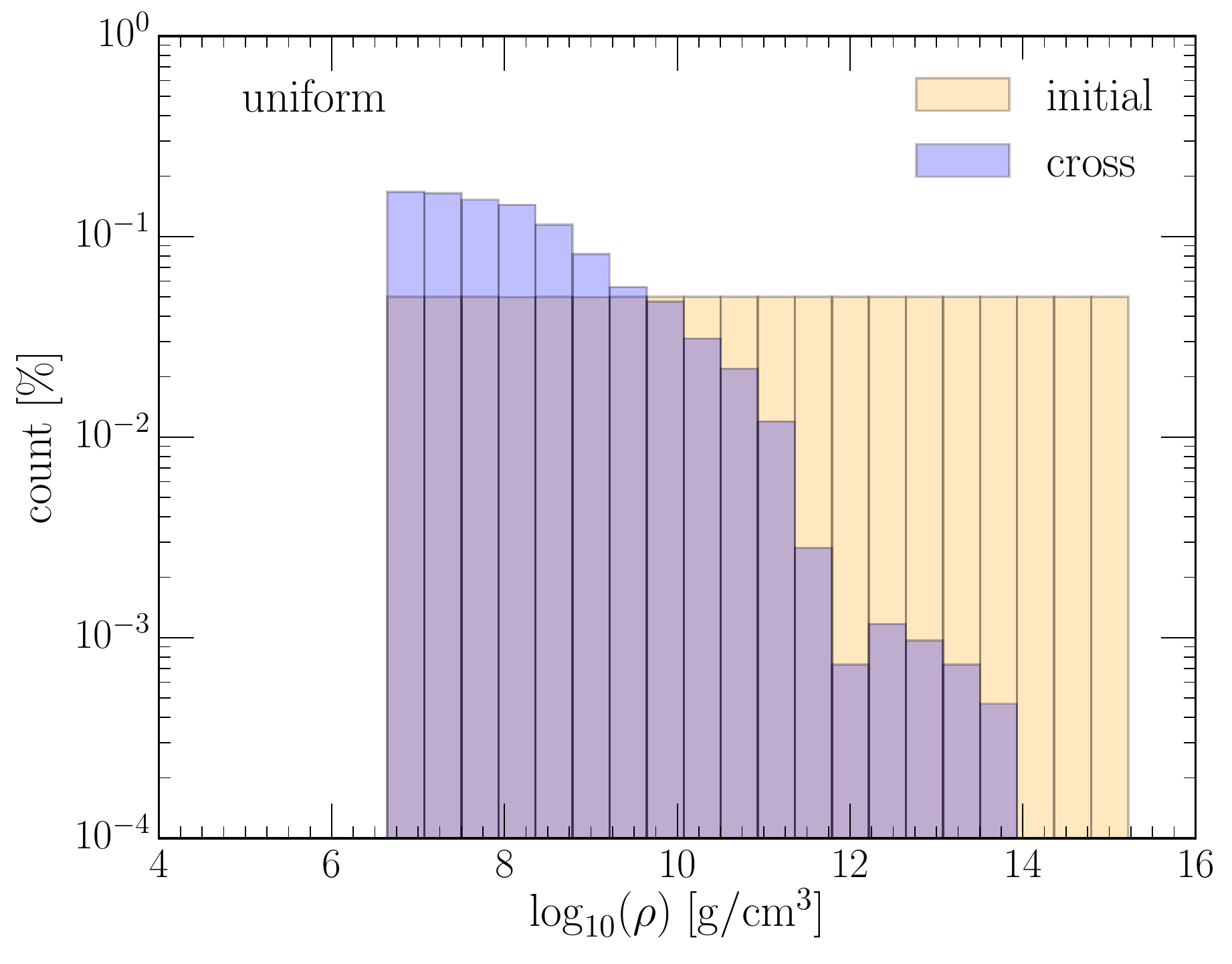}
\includegraphics[width=0.5\columnwidth]{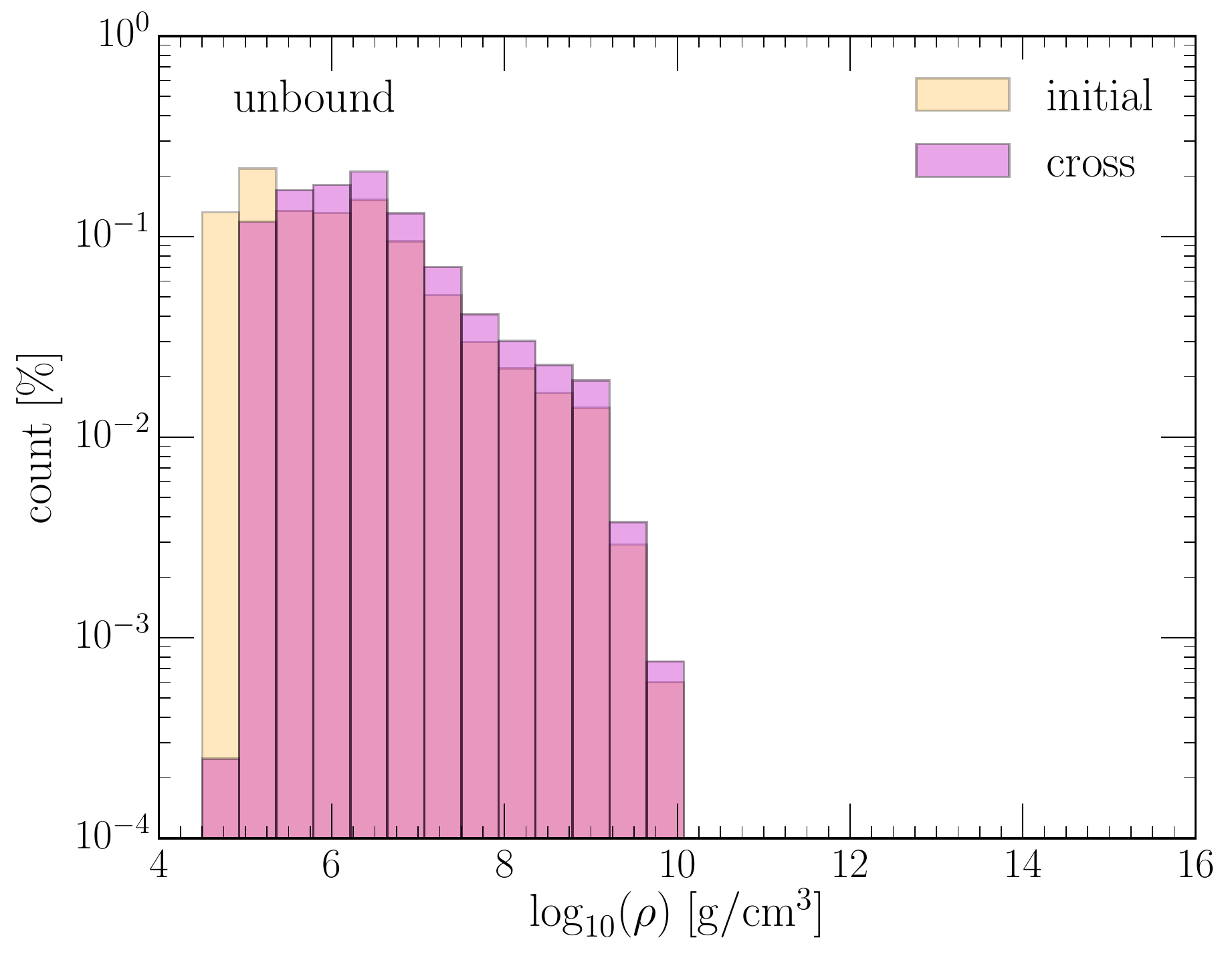}
\caption{Distributions in rest-mass density of the tracers (light-yellow
  shade) and the corresponding distributions when the tracers cross the
  $200\,\Msun$ surface. Shown from left to right are the distributions
  corresponding to the ``correlated'', ``anticorrelated'', ``uniform'',
  and ``unbound''.}
  \label{fig:init_comparison}
\end{figure*}

This is a somewhat surprising result as one would have naively expected
that the ``correlated'' placement would have provided a more faithful
representation of the fluid dynamics. In order to appreciate why this is
not the case, we report in Fig.~\ref{fig:init_comparison} the initial
distributions in rest-mass density of the tracers (light-yellow shade)
and the corresponding distributions when the tracers cross the
$200\,\Msun$ surface. In particular, the comparison presented in the top
left panel allows us to appreciate that initially there were
significantly more tracers at low rest-mass densities for the
``correlated'' placement, so that we are effectively oversampling at
lower rest-mass densities and, consequently, undersampling at high
rest-mass densities. Indeed, there are no tracers that come from above
$\sim 10^{12}\gcm$ while almost all tracers below $10^{9} \gcm$ become
unbound. Because the lower densities also have higher values of $Y_{e}$,
the ``correlated'' placement de-facto leads to an oversampling around
$Y_{e} \sim 0.18$ (hence the second broad peak in the distribution) and
to an undersampling at $Y_{e}\sim 0.05$ (hence the smaller first peak in
the distribution). Not surprisingly, the ``anticorrelated'' placement
suffers of the opposite problem. It undersamples at lower rest-mass
densities resulting in there being fewer tracers with large values of
$Y_{e}$. More precisely, the top right panel of
Fig.~\ref{fig:init_comparison} shows that there is a significant number of
tracers above $10^{12} \gcm$, which have low values of $Y_{e}$, thus
yield a good agreement with the first peak. However, at lower rest-mass
densities, a higher percentage of tracers that become unbound come from
lower densities and hence we are effectively and significantly
undersampling matter at low densities. For example, only about $1\,\%$ of
the initial tracers were placed at the lowest initial rest-mass density
of $\sim10^{7} \gcm$, but almost $20\,\%$ of the tracers that are unbound
come from these rest-mass densities. Examining the bottom left panel of
Fig.~\ref{fig:init_comparison} we can also appreciate that the
``uniform'' placement distribution is, in a sense, the average of the
above two placements. It uniformly samples from all densities and hence
has no oversampling or undersampling at high and low rest-mass
densities. As shown in Fig.~\ref{fig:init_comparison} tracers that are
initially at $10^{14}\gcm$ become unbound, as in the case of the
``anticorrelated'' placement; at the same time, the tracers at low
rest-mass densities are unbound, thus making the ``uniform'' placement
be the most effective one, besides being also the simplest
to implement.

\begin{figure*}
\includegraphics[width=0.33\columnwidth]{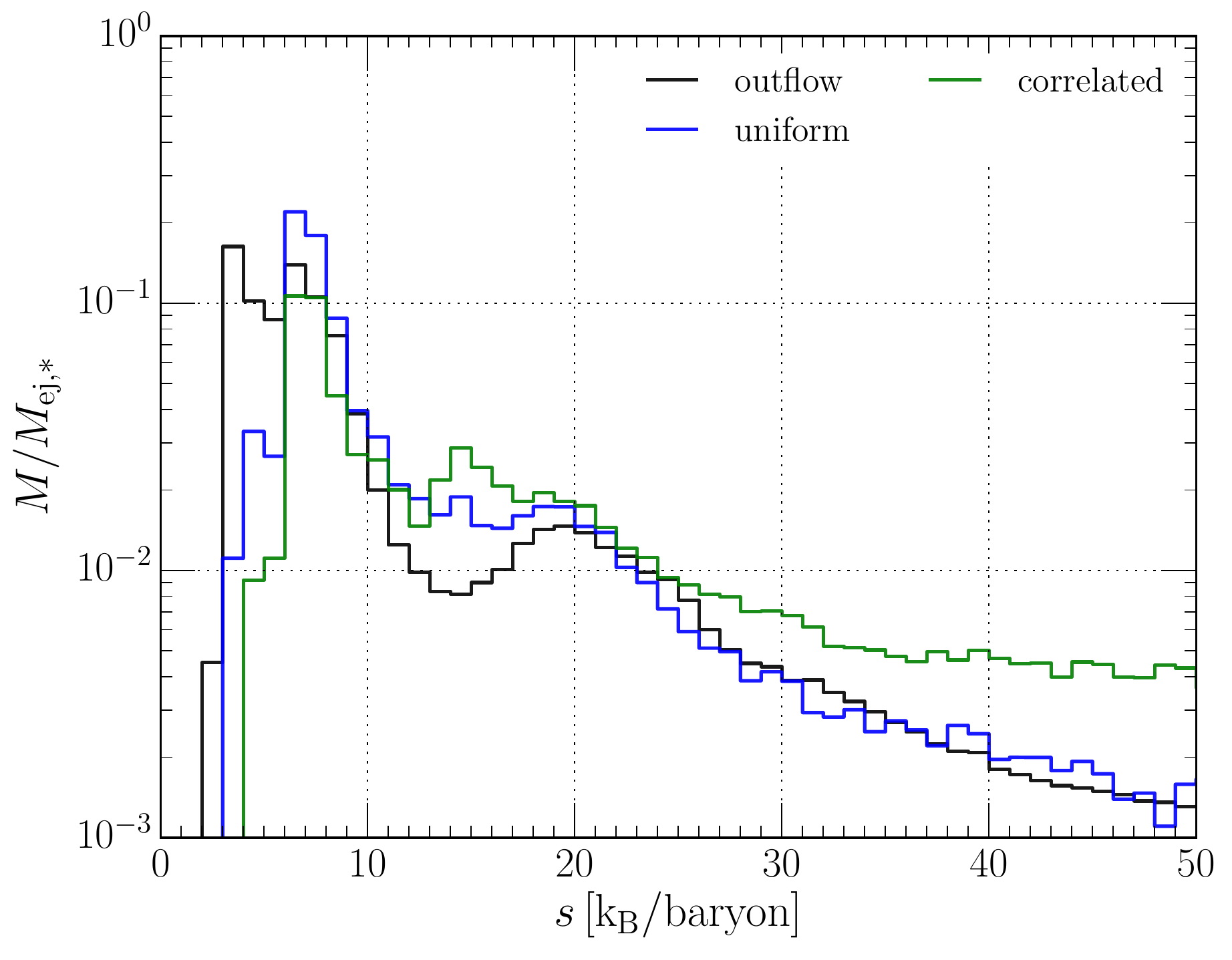}
\includegraphics[width=0.33\columnwidth]{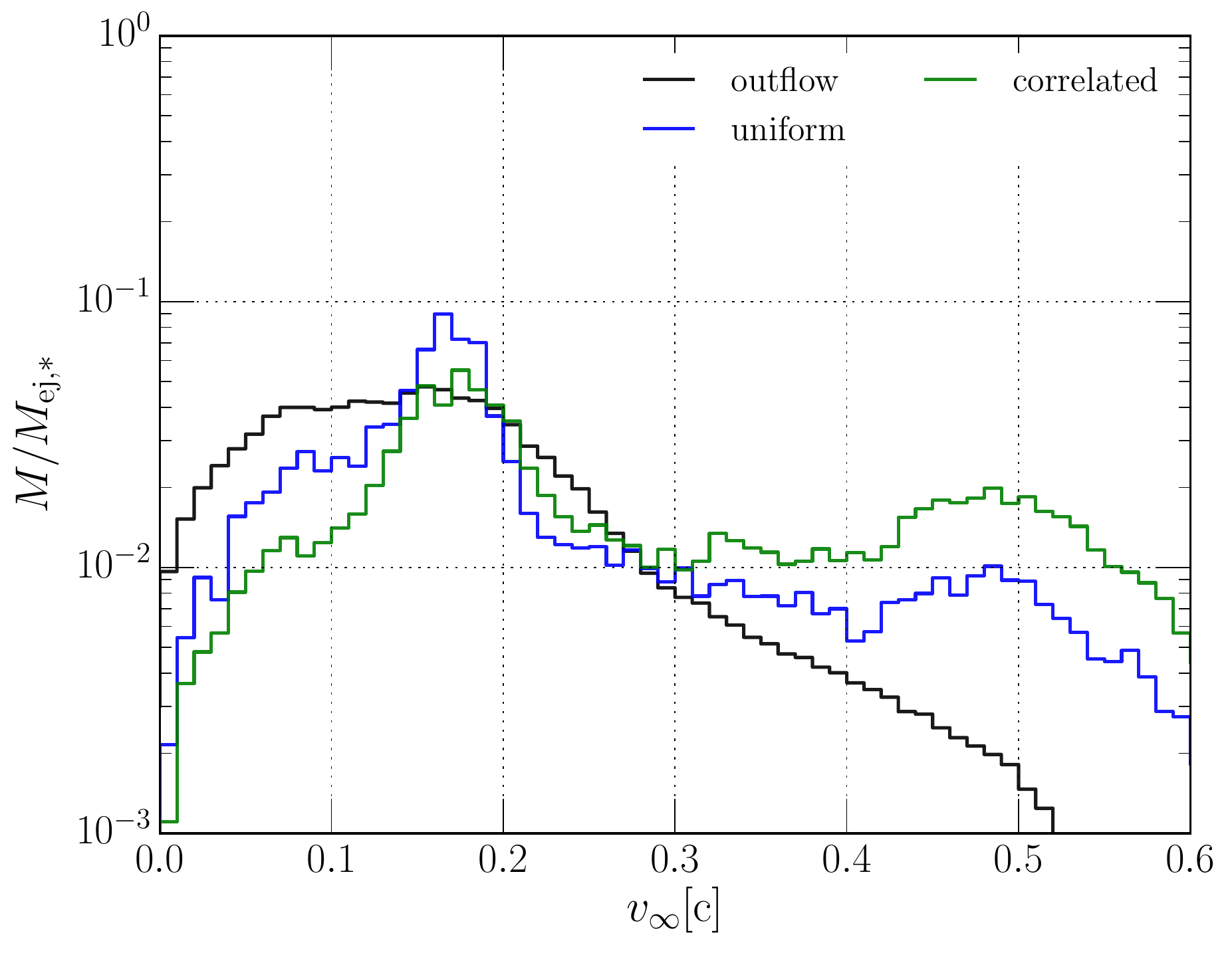}
\includegraphics[width=0.33\columnwidth]{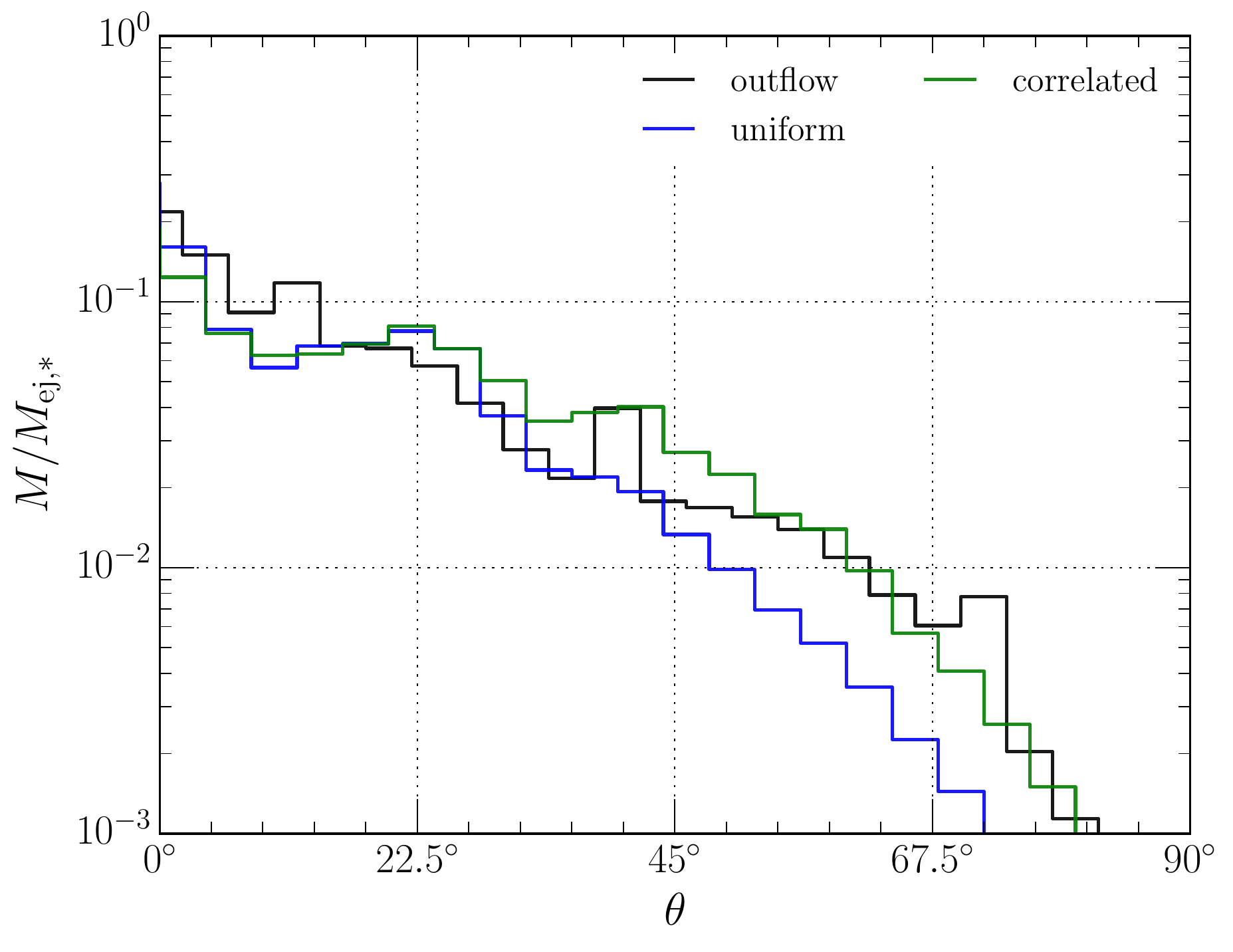}
\caption{Comparison between the distributions in electron fraction as
  computed with the tracer mass-fluxes through the surface at
  $200\,\Msun$ as computed with the different placement schemes (coloured
  lines) and the outflow (black line). The quantities considered are: the
  entropy, the velocity at infinity, and the angular distribution; all
  curves have been normalised to the corresponding total ejected masses
  (\cf Fig.~\ref{fig:comp_methods}).}
  \label{fig:comp_methods_alt_props}
\end{figure*}

When considering the ``unbound'' placement, the bottom right panel of
Fig.~\ref{fig:init_comparison} shows that there is a remarkably good
match between the initial distribution of the placed tracers and the
distribution at detector crossing, especially for medium and high
rest-mass densities. Yet, while such a placement nicely reproduces the
counts at the two peaks in electron fraction, it also fails to capture
the values between the two peaks, as shown in the left panel of
Fig.~\ref{fig:comp_methods}. This is most likely due to the fact that
this placement completely misses some of the material that the other
tracer placements instead capture. In particular, it undersamples the
material that is at low rest-mass density and that can become unbound as
a result of shock heating. Hence, the behaviour in
Figs.~\ref{fig:comp_methods} and \ref{fig:init_comparison} highlights a
potential drawback of the ``unbound'' placement procedure, that one may
naively assume to be particularly good. This placement, in fact, while it
can nicely sample the unbound material, only captures material that is
unbound at a given moment and neglects any material that might become
unbound and that has not yet become unbound, most notably, the very low
rest-mass density material.

Finally, in Fig.~\ref{fig:comp_methods_alt_props} we compare other
thermodynamical and physical quantities similar the one presented in Fig.
\ref{fig:comp_methods}. These three quantities are: the specific entropy
(left), the velocity at infinity (middle), and the angular distribution
(right).  Here, we concentrate on a comparison between the fluid outflow
(black line), the ``uniform'' distribution placement (blue line), and the
``correlated'' placement (green line). Note how the specific-entropy
distributions are all rather similar, although the ``uniform'' placement
undersamples at the first peak and does not show a local minimum at
around $15\,{\rm k_{B}/baryon}$. The velocity at infinity is another
measure of the $u_{t}$ and here we do not see great agreement at higher
values. These higher values correspond to the more energetic tracers,
thus indicating that both the ``uniform'' and the ``correlated''
placement tend to oversample the highly energetic material. Similar
considerations apply to the angular distribution of the ejected, where
the two placement schemes show rather similar results, both in agreement
with the fluid evolution, even though the tracer mass flux shows slightly
more mass at the higher angles. Interestingly, the tracer evolution is
also able to reproduce the local maximum at $\sim 45^{\circ}$ which a
numerical artefact of the Cartesian grid and was first mentioned in
Ref.~\cite{Radice2016}.

\medskip
\emph{In conclusion, we have found that out of the initial tracer
placements considered, the ``uniform'' placement results in the best
agreement with the underlying fluid based on a comparison with the
electron fraction. This is due to the ``uniform'' method sampling both
high and low rest-mass densities uniformly, avoiding over- or
undersampling in these regimes.}

\section{Tracing bound material}
\label{sec:bound}

We conclude our discussion on the use of tracers by considering the case
in which they are employed to describe the dynamics of matter that is
gravitationally \emph{bound}. Despite already having a complete
description of the fluid dynamics as measured by Eulerian observers,
there are a number of reasons why using tracers could be a powerful tool.
The most important of such reasons is that tracers following bound fluid
elements allow us to disentangle the local dynamics of the fluid from the
global one. Because both of these dynamics can be complex and operate on
different timescales, having the possibility of setting them apart is
quite valuable to interpret the results of simulations. There are
several examples of the successful use of tracers in bound material (see,
\eg the interesting analysis of precessing and tilted disk accretion in
Ref.~\cite{Mewes2016}), but we will here concentrate on the dynamics of
the HMNS produced by the merger of a binary system of neutron stars,
following the discussion already presented in Ref.~\cite{Hanauske2016}. 

More specifically, we are interested here in studying the rotational
properties of the HMNS produced by the merger of a binary system of
neutron stars. The dynamics of this object, which is metastable and
doomed to collapse to a black hole because its mass is above that
sustainable through uniform rotation, is made complicated by the fact
that it is subject to violent oscillations and, at the same time, it is
subject to a barmode deformation that leads to a copious emission of
gravitational waves. Hence, a deeper understanding of the distribution of
angular momentum and angular velocity can help in determining under what
conditions the HMNS will collapse to a black hole. It is therefore useful
to study the angular velocity of the fluid, which we measure as
\begin{equation}
\Omega := \alpha\, v^\phi - \beta^\phi \,,
\label{eq:omega2}
\end{equation}
where $\phi$ refers to the azimuthal direction (see \cite{Hanauske2016}
for a more detailed discussion). Hereafter we will concentrate on the
HMNS produced when two identical neutron stars with a (gravitational)
mass of $1.35\,\Msun$ and described by the hot DD2 EOS, but a much large
sample of binaries has been explored in Ref.~\cite{Hanauske2016}.

\subsection{Initial placement}

Selecting the optimal initial placement when studying bound flow is much
less problematic since we are not trying to accurately capture the
statistics of the underlying fluid. Indeed, so long as it is clear as to
what region of the fluid we are interested in, then placing tracers is
mostly a matter of deciding a good time for the seeding and employing a
sufficiently large number of tracers so that all parts of the flow are
properly represented. Because we are interested in tracking the
motion in the core of the HMNS, it is straightforward to set the time of
the merger as the time for the tracer seeding. Furthermore, for the
spatial placement we do not follow any probability distribution
associated with the rest-mass density but simply place a single tracer in
each of the fluid cells that we mark to be within a certain region of the
computational domain, namely, the core of the HMNS. In practice, using
$\sim 10^5$ tracers is sufficient to ensure that every cell within the
core had a tracer in it. As in Ref.~\cite{Hanauske2016}, of the $10^5$
initially placed tracers, approximately $6000$ remained on the $(x,y)$
plane, thus allowing us to study the fine details of the fluid motion on
this plane.

\subsection{Dynamics of the HMNS}

\begin{figure}
\includegraphics[width=\columnwidth]{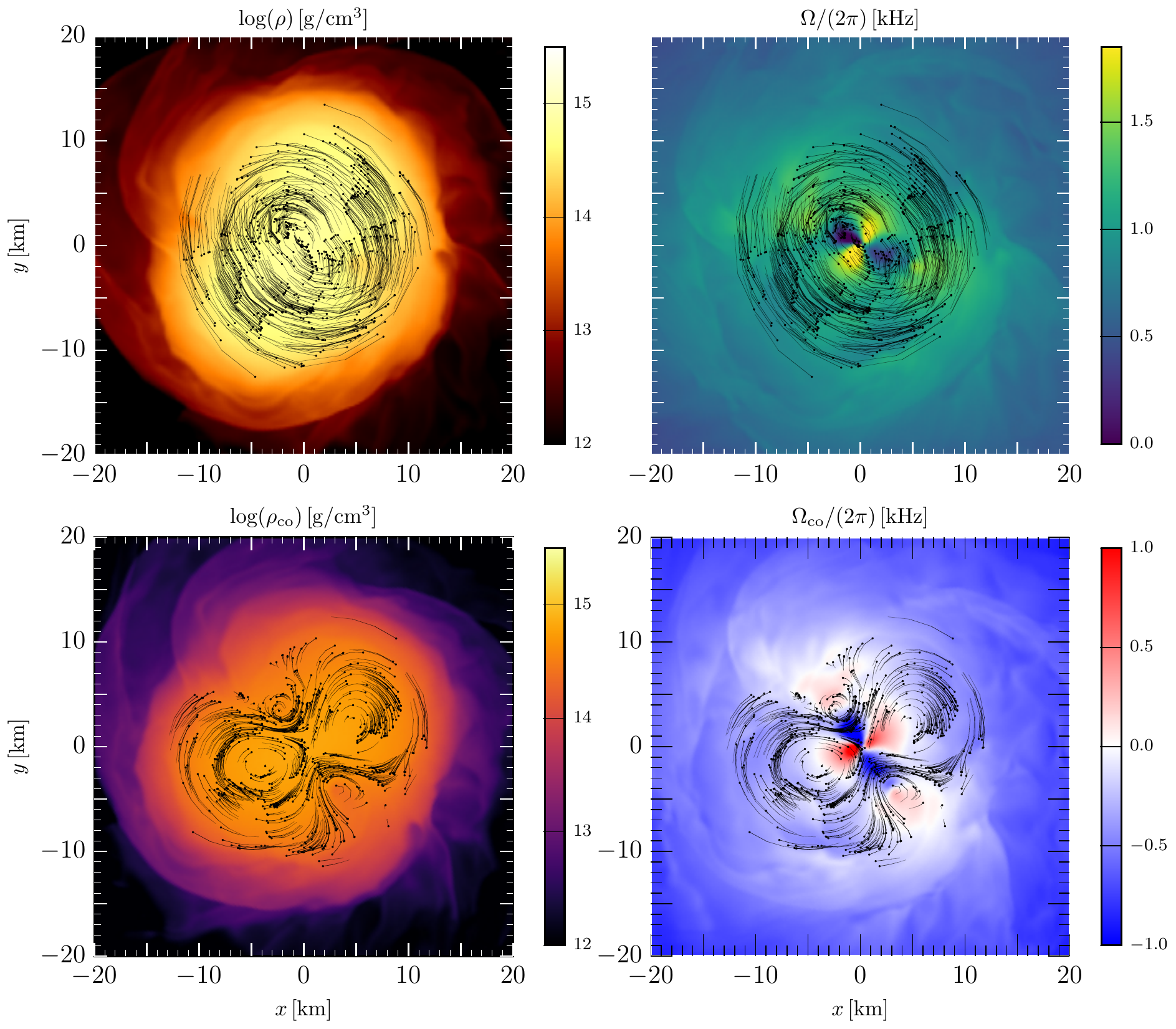}
\caption{Distributions on the $(x,y)$ plane of the rest-mass density
  (left panels), and of the angular velocity (right panels). The top
panels are relative to the Eulerian frame at time $t=5.5\,\mss$, while
the bottom panels refer to a corotating frame with angular velocity
$\Omega_{\rm co} := \Omega - \Omega_{\rm GW}/2$, where $\Omega_{\rm
GW}=2582 \,{\rm Hz}$ is the instantaneous gravitational-wave frequency.
Also shown are portions of the flowlines of several tracer particles that
remain close to the $(x,y)$ plane and for which we show only the final
part of the flowlines (\ie for the last $\simeq 0.285\mss$), using small
dots to indicate the particle position at the time indicated in the
frame. In addition, the initial parts of the trajectories have increasing
transparency so as to highlight the final part of the trajectories.}
\label{fig:core_evo}
\end{figure}

In Fig.~\ref{fig:core_evo} we plot a representative time, approximately
$\sim5.5\mss$ after the merger, of the evolution of the rest-mass density
(left) and the angular velocity (right). In the top panel we chose the
frame of an observer at infinity (Eulerian), while in the bottom we chose
the ``corotating frame''\footnote{We recall that the corotating frame is
  a frame rotating at a frequency that is half of the instantaneous
  gravitational-wave frequency, $\Omega_{\rm GW}$. Here $\Omega_{\rm
    GW}/2 = 1291 \,{\rm Hz}$.}. The tracers plotted here are just a
representative of the $\sim 6000$ tracers that remain in the $(x,y)$
plane throughout the entire evolution. For each tracer we use a small
filled circle to report the position at the given time, and use solid
lines of decreasing size the join the position at two previous times
separated by $\sim 0.3\,\mss$. The ``tails'' that are produced in this
manner provide a simple way to visualise the tracer streamlines and hence
the evolution of the fluid as it rotates around the core. Also, the
number of tracers is sufficient to completely cover the core and all key
features are represented by numerous tracers, however only representative
ones are plotted to illustrate the rich evolution of the feature that is
difficult to observe without the streamlines.

The tracers immediately clarify and illustrate the evolution of the
fluid. As seen in the bottom left panel of Fig.~\ref{fig:core_evo}, the
underlying rest-mass density clearly exhibits an $l=m=2$ deformation;
however, because this deformation is not stationary in an Eulerian frame,
it is difficult to discern the motion of the fluid, as can be observed by
the tracers in the top left panel. With tracers, on the
other hand, we can clearly see that there is a central bar around which
the tracers flow, but beside this bar there are two vortices that trap
fluid (bottom of Fig.~\ref{fig:core_evo}). This pattern can be more
clearly seen in the angular velocity, where these side vortices have
higher angular velocity compared with the central bar. Additionally, the
tracers illustrate further the structure of the angular velocity
distribution. In particular, in the vortices the fluid is revolving
faster than the corotating frequency -- and hence move counterclockwise
-- while other tracers either corotate -- hence they appear not to move
-- or rotate more slowly than the corotation frequency -- and hence move
clockwise. The combined use of tracers and the corotating frame
demonstrate the ability to visualise features of the core that are
otherwise difficult to visualise.

\begin{figure}
\includegraphics[width=\columnwidth]{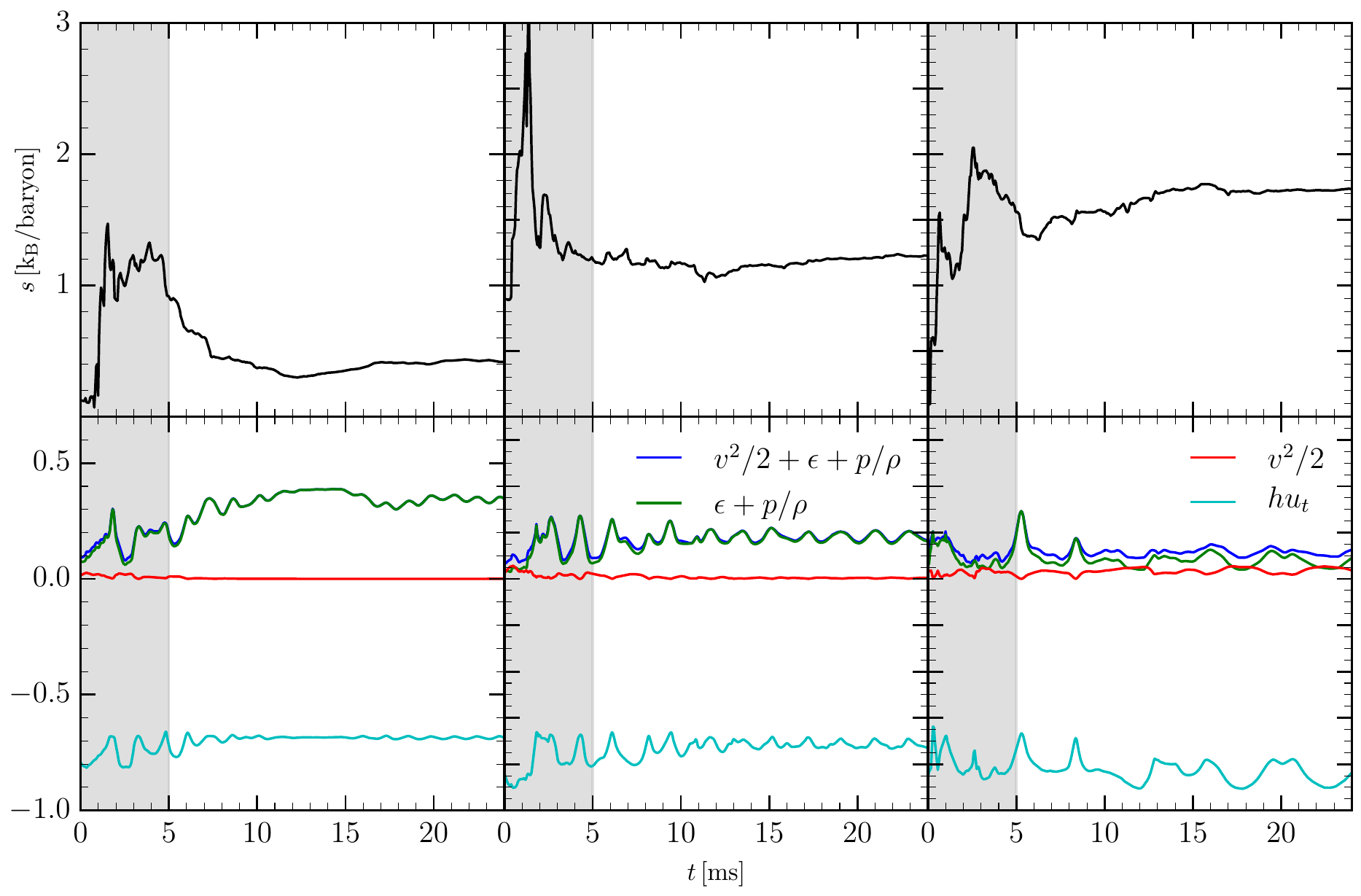}
\caption{\textit{Top panels:} Evolution of the entropy for tracers in the
  HMNS that are eventually either in the inner core (\ie $r \lesssim
  4\,\km$; left panel), in the middle core (\ie $4 \lesssim r \lesssim
  6\,\km$; middle panel), or in the outer core (\ie $6 \lesssim r
  \lesssim 8\,\km$; right panel). \textit{Bottom panels:} evolution of
  the Bernoulli constant $hu_{t}$ (light blue line), the classical
  Bernoulli constant $v^{2}/2+\epsilon + p/\rho$ (blue line) and its main
  contributions: $\epsilon+p/\rho$ (green line) and the kinetic term
  $v^{2}/2$ (red line). The grey-shaded area refers to the postmerger
  transient when the HMNS is far from an equilibrium.}
\label{fig:bernoulli}
\end{figure}

Another advantage of the tracers is that they allow the computation of
quantities along their streamlines and, in particular, of quantities that
are expected to be conserved if the fluid is perfect and the flow is
isentropic, \eg the Bernoulli constant. Keeping track of such quantities
can be extremely useful to explain a behaviour which would otherwise
appear puzzling. For instance, in Ref.~\cite{Hanauske2016} it was
proposed that the dependence of the angular velocity depends inversely on
the rest-mass density, through the Bernoulli constant. This conjecture
explains why the rest-mass density and angular velocity profiles have a
clear $90$ degree phase difference (\cf left and right panels of
Fig.~\ref{fig:core_evo}). This assertion, however, is difficult to verify
without tracers and the streamlines they can be provide.

We recall that in relativistic hydrodynamics and for a perfect fluid with
four-velocity $\boldsymbol{u}$, the quantity
$h\,(\boldsymbol{u}\cdot\boldsymbol{\xi})$ is Lie-dragged along
$\boldsymbol{u}$ \cite{Rezzolla_book:2013}
\begin{equation}
\label{eq:bern0}
L_{\boldsymbol{u}}
(h\,\boldsymbol{u}\cdot\boldsymbol{\xi})=0 \,,
\end{equation}
where $h:= (e+p)/\rho$ is the specific enthalpy, $e:=\rho(1+\epsilon)$ is
the total energy density, and $\boldsymbol{\xi}$ is a Killing vector of
the spacetime and also a generator of the symmetry obeyed by the
fluid. If spacetime admits a timelike Killing vector so that the fluid
motion is stationary, then the quantity $\mathcal{B}:= h u_t$ is a
constant of the fluid. In its classical limit, Eq.~(\ref{eq:bern0})
becomes
\begin{equation}
\label{eq:bern1}
\left(1+\epsilon+\frac{p}{\rho}\right) \left( 1 + \phi +
\frac{1}{2}\vec{\boldsymbol{v}}^2 \right) =
{\rm const.}\,,
\end{equation}
where $\phi$ is the gravitational potential and $\boldsymbol{v}$ is the
local fluid velocity. When neglecting higher-order terms and taking the
gravitational potential to be independent of time and essentially
constant across the HMNS, expression (\ref{eq:bern1}) further reduces to
\begin{equation}
\label{eq:bern2}
\left(
\frac{1}{2}\vec{\boldsymbol{v}}^2 + \epsilon + \frac{p}{\rho}
\right) =
{\rm const.}\,,
\end{equation}
which coincides with the classical expression for the Bernoulli constant
\cite{Rezzolla_book:2013}. Clearly, the post-merger phase of a BNS merger
does not satisfy the timelike Killing vector assumption, but we can still
use tracer particles to appreciate how well the Bernoulli constant is
conserved along streamlines.

Figure~\ref{fig:bernoulli} provides a simple example of how the evolution
of tracers allows one to quickly and simply perform the analysis of very
precise regions of the fluid. In particular, the panels in
Fig.~\ref{fig:bernoulli} show the time evolution of several quantities
relative to three tracers that have been chosen as representative of
three areas areas of the HMNS: the inner core (\ie $r \lesssim 4\,\km$),
the middle core (\ie $4 \lesssim r \lesssim 6\,\km$), and the outer core
(\ie $6 \lesssim r \lesssim 8\,\km$). The top panels, in particular, show
the evolution of the specific entropy which, during the violent
post-merger phase, undergoes a sudden jump due to heating, but quickly
settles to become approximately constant. The bottom panels, instead,
shows the evolution of the Bernoulli constant in light blue. 

In essence, Fig.~\ref{fig:bernoulli} shows that even though the
post-merger phase does not exhibit a timelike Killing vector, the
Bernoulli constant stays roughly constant across the interior of the
HMNS. This constant behaviour can be used to explain the phase difference
observed in Ref.~\cite{Hanauske2016} where the rotation-profile is
out-of-phase with the barmode from the rest-mass density. In fact, while
the values of (\ref{eq:bern2}) are strictly not constant in time
(especially in the grey-shaded areas), they also do not vary
significantly around the initial values. More importantly, there is a
phase opposition in the evolution of the kinetic term $v^2/2$ and of the
pressure term $\epsilon + p/\rho$, so that large values of the former
correspond to low values of the latter and vice versa, thus providing a
simple and intuitive explanation for the phase difference in the two
quantities.

\section{Conclusions}
\label{sec:conclusions}

Accurate numerical simulations of the merger of binary systems of neutron
stars are becoming increasingly more important as the detection of
gravitational waves from such systems is expected within the next few
years and is likely to bear a wealth of information in physics and
astrophysics. A particularly important aspect of this process that has
recently received considerable attention, is the ejection of matter from
the merger, which is expected to be well suited for r-process
nucleosynthesis \cite{Lattimer74, Li1998, Tanvir2013, Berger2013,
Tanaka2013, Rosswog2014a, Sekiguchi2015, Radice2016}. Because it
naturally produces neutron-rich material, the merger of BNSs could
represent the astrophysical site behind the origin of the heavy elements
in our universe. A number of numerical simulations, in fact, have
recently demonstrated that BNS mergers can provide a contribution to the
chemical abundances of heavy elements that is several orders of magnitude
larger than that given by supernovae, whose ejected material is not
enough neutron-rich to match with observed values in the solar system. In
addition, this ejected r-process material will likely produce an
electromagnetic counterpart known as a kilonova that can additionally
provide a wealth of information about the nuclear physics of the
r-process.

However, due to the long timescales required for the nuclear-reaction
networks, present numerical-relativity simulations are unable to take
directly into account the complex nuclear reactions, which must therefore
be analysed in a post-processing step. It follows that there is a need
for robust and accurate methods that can be combined with the
nuclear-reaction networks to produce abundance patterns from merger
simulations. Tracers, that is, massless particles that are passively
advected, are an effective way of recording the thermodynamical history
of the fluid in grid-based codes and as such have been employed by a
number of recent simulations of merging neutron-star binaries. However,
while not particularly expensive to compute, the use of tracers requires
a judicious initial placement, both in time and space, and especially if
the portion of the fluid one is interested in is the one that is
gravitationally unbound. This is because different placements can
potentially lead to different physical observables and hence to different
physical conclusions on the properties of the underlying fluid.

In this paper we have investigated the efficacy of how well the tracers
can match the underlying fluid by directly comparing results from the
fluid itself. More specifically, we have evaluated four different
placement schemes that could be used to initially distribute the tracers
and how well their predictions match those obtained when using
information from the actual fluid flow. The four schemes considered are
built by constructing distribution functions of the tracers that are
either directly \emph{correlated} with the rest-mass density
distribution, or \emph{anti-correlated}, or \emph{uniformly} distributed,
or, as a final case, independent on the rest-mass density and related
instead to the portion of fluid that is \emph{unbound}. Countering our
naive expectations that placements that are correlated with the rest-mass
density distribution or that follow the fluid that is marked unbound, we
have found that the most effective method is to \emph{uniformly} sample
across the rest-mass density distribution as this leads to the closest
matching with the unbound material that is flowing through a
detector. This match has been measured in terms of a large number of
physical quantities, such as the unbound rest-mass, the distributions of
electron fraction, specific entropy and kinetic energy. Interestingly,
the most important reason why a uniform distribution represents the
optimal one is that it has the merit of sampling sufficiently from both
the high and low rest-mass density regions that are ejected during the
merger. Other methods, in fact, tend to either underestimate or
overestimate the contributions coming form certain regions of the fluid.

In addition, tracers can be also used for a fine analysis of the motion
of the fluid in regions that are highly complex, providing information on
the properties of the fluid and on the quantities that should be
conserved along streamlines if the flow is adiabatic. We have illustrated
this by considering the motion of tracers in the core of an HMNS produced
in the merger of a BNS system. We have shown that tracers provide a novel
way to show the evolution of the fluid and display features that are
otherwise difficult to observe and analyse. As a representative example,
we have shown that the fluid vortices can form in the HMNS where matter
is locally trapped and that these vortices are located in regions of low
rest-mass density and high fluid velocity, as expected from the
conservation of the Bernoulli constant. In a forthcoming work, we use the
results discussed here to investigate a wide range of EOS and masses and
the nucleosynthesis produced from the tracers ejected during the
dynamical-ejection phase. Additionally, we will use the tracer data as
input for radiative-transfer models to calculate the light curves
resulting from the corresponding kilonova \cite{Bovard2017}.

\section*{Acknowledgements}
It is a pleasure to thank Filippo Galeazzi, Matthias Hanauske,  Bruno
Mundim, David Radice for useful discussions. This research is supported
in part by the ERC synergy grant ``BlackHoleCam: Imaging the Event Horizon
of Black Holes" (Grant No. 610058), by ``NewCompStar'', COST Action
MP1304, by the LOEWE-Program in the Helmholtz International Center (HIC)
for FAIR, and by the European Union's Horizon 2020 Research and
Innovation Programme (Grant 671698, call FETHPC-1-2014, project
ExaHyPE). LB is supported by HIC for FAIR and the graduate school
HGS-HIRe. The simulations were performed on the SuperMUC cluster at the
LRZ in Garching, on the LOEWE cluster in CSC in Frankfurt, and on the
HazelHen cluster at the HLRS in
Stuttgart.
\bigskip
\bigskip

\providecommand{\newblock}{}

\bibliographystyle{iopart-num}

\end{document}